\documentclass[12pt,aps, prd, reprint,preprintnumbers,onecolumn,nofootinbib]{revtex4-2}
\usepackage[most]{tcolorbox}
\usepackage{graphicx}
\usepackage{amsmath}
\usepackage{amsfonts}
\usepackage{amssymb}
\usepackage{hyperref}
\usepackage{soul}
\usepackage{xcolor}
\usepackage{cancel}
\usepackage{physics}
\usepackage{cleveref}
\usepackage{orcidlink}
\usepackage[top=1in, bottom=1in, left=1in, right=1in]{geometry}

\newcommand\redout{\bgroup\markoverwith
{\textcolor{red}{\rule[.5ex]{2pt}{0.4pt}}}\ULon}

\begin{document}

\preprint{LA-UR-23-34053}
\preprint{IQuS@UW-21-078}

\title{Three-flavor Collective Neutrino Oscillations on D-Wave's {\tt Advantage} Quantum Annealer } 


\author{Ivan A. Chernyshev,\orcidlink{0000-0001-8289-1991} } 
\email{ivanc3@uw.edu}
\affiliation{InQubator for Quantum Simulation (IQuS), Department of Physics, University of Washington, Seattle, WA 98195, USA.}
\affiliation{Theoretical Division, Los Alamos National Laboratory, Los Alamos, NM 87545, USA.}



\date{\today}

\begin{abstract}
In extreme environments such as core-collapse supernovae, neutron-star mergers, and the early Universe, neutrinos are dense enough that their self-interactions significantly affect, if not dominate, their flavor dynamics. In order to develop techniques for characterizing the resulting quantum entanglement, I present the results of simulations of Dirac neutrino-neutrino interactions that include all three physical neutrino flavors and were performed on D-Wave Inc.'s {\tt Advantage} 5000+ qubit quantum annealer. These results are checked against those from exact classical simulations, which are also used to compare the Dirac neutrino-neutrino interactions to neutrino-antineutrino and Majorana neutrino-neutrino interactions. The D-Wave {\tt Advantage} annealer is shown to be able to reproduce time evolution with the precision of a classical machine for small numbers of neutrinos and to do so without Trotter errors. However, it suffers from poor scaling in qubit-count with the number of neutrinos. 
\end{abstract}


\maketitle


\section{Introduction}
\label{sec:intro}

In extreme-density and extreme-temperature environments, such as core-collapse supernovae (CCSNe), neutron-star mergers and the early Universe, neutrinos dominate the transport of energy, momentum, entropy, neutron-to-proton ratios, and lepton flavor composition, among other characteristics (for recent reviews, see Refs. \cite{janka2007theory, Mezzacappa_2020, burrows2021corecollapse, fuller2023neutrinos, foucart2023neutrino}). Several neutron star merger studies show neutrinos having an effect in phenomena such as mass ejection and accretion, as well as gamma-ray burst formation \cite{ruffert_coalescing_1997, radice_long-lived_2018, nedora_numerical_2021}. 
In CCSNe, neutrinos carry away 99\% of the gravitational binding energy released by the collapse of the iron core at the supernova's beginning \cite{1982ApJ263366B, woosley1988supernova, turatto1998peculiar, sollerman1998very, nicholl2020extremely, smith2007sn, drake2010discovery, chatzopoulos2011sn, rest2011pushing, benetti2014supernova, pagliaroli2009improved}. There is also a consensus in the field that these neutrinos are the primary triggering mechanism in most CCSNe \cite{colgate1966hydrodynamic, arnett1966gravitational, 1982ApJ263366B, wilson_supernovae_1985, bethe1985revival, fischer_neutrino_2012, fischer_protoneutron_2010, huedepohl_neutrino_2010}. 
Neutrinos are also thought to play a key role in nucleosynthesis in all three of the above phenomena, with a few light isotopes and several heavy ones being particularly dependent on neutrino processes \cite{savage1991neutrino, qian_connection_1993, frohlich2006neutrino, grohs_surprising_2016, just_comprehensive_2015, wanajo_rp-process_2006, woosley_nu-process_1990, roberts_integrated_2010, grohs_surprising_2016}. 
One challenge is that at such high densities and temperatures, neutrino-neutrino interactions, which manifest as collective neutrino oscillations, are significant enough to produce macroscopic effects. 
Recent simulations have shown that a few neutrino-dependent processes which are theorized to be critical in the nucleosynthesis of nuclei with mass numbers above 64 are heavily dependent on collective neutrino oscillations \cite{frohlich2006neutrino, sasaki2017possible, balantekin2023collective}. 
Collective neutrino oscillations are also expected to have an effect on r-process, which is responsible for half of the abundance of elements in the Universe heavier than iron \cite{duan2011influence}. In short, collective neutrino oscillations must be taken into account in order to fully understand the dynamics of neutrinos in high-temperature, high-density astrophysical phenomena.

Collective neutrino oscillations are a highly nonlinear many-body problem \cite{fuller_resonant_1987, notzold_neutrino_1988, sigl_general_1993}. One commonly-used method for highly nonlinear many-body problems is mean-field theory, which treats all but the observed neutrino as a single-valued background interacting with said neutrino \cite{qian_neutrino-neutrino_1995}. This approach is effective in achieving a linear scaling of computational resources with the number of modes in the neutrino-state-space considered in the simulation. This is typically sufficient for studies at the astrophysical scale to take place. However, it is unable to simulate quantum entanglement and incoherent scattering. Whether this fact invalidates the use of mean-field theory in macroscopic neutrino systems is still an unanswered question \cite{shalgar2023evidence}. 


Meanwhile, methods that do account for entanglement in collective neutrino oscillations have been utilized extensively. They include exact time evolution through numerical integration or diagonalization of the Hamiltonian or the equations of motion, the Bethe ansatz, and tensor network algorithms \cite{Bethe1931Metalle, vidal2003efficient, pehlivan2011invariants, Espinoza2013, Pehlivan2014, birol2018neutrino, Cervia2019, Cervia2019Entanglement, patwardhan2019eigenvalues, rrapaj2020exact, patwardhan2021spectral, roggero2021dynamical, roggero2021entanglement, cervia2022collective, Lacroix2022, Martin2023ManyBody,  Fiorillo2023a, martin2023equilibration, cirigliano2024neutrino, Bhaskar2024TimeScales, neill2024scattering}. 
For neutrino systems with geometrically symmetric and simple velocity distributions, generalized angular momentum representations can be used to conduct simulations at scale \cite{friedland2003manyparticle, friedland2006construction, xiong2022manybody, roggero2022entanglement, Martin2022Classical}. However, as is the case with many simulations of quantum-mechanical phenomena, the resources needed to study entanglement in the general case of collective neutrino oscillations on classical computers scales exponentially with the size of the system \cite{balantekin2023quantum, hall2021simulation, balantekin2023collective, patwardhan2023, roggero2021entanglement, siwach2023entanglement}. As a result, simulations of collective neutrino oscillations in the using exact numerical methods have been limited to up to 20 neutrinos. 
Tensor network \cite{roggero2021dynamical, roggero2021entanglement, cervia2022collective} and matrix product state methods \cite{vidal2003efficient, schollwock2011densitymatrix, paeckel2019timeevolution}, which are two of the leading methods of approximate simulation of quantum mechanical systems, can simulate considerably more neutrinos. 
For instance, the time-dependent variational principle (TDVP) can currently be used to simulate up to approximately 50 neutrinos if the neutrinos start in the same flavor, though performance has been found to be little better at scale than exact simulations for a mix of initial flavors \cite{cervia2022collective}.

In the end,  the fundamental problem of exponential scaling of computational resource requirements with system size has not been solved for simulations that utilize state-of-the-art computers. This presents an opportunity for quantum computers to produce an advantage, as they are capable of efficiently representing any local quantum system \cite{feynman_1982, lloyd_1996}. Simulations of two-flavor collective neutrino oscillations have been done on devices such as Quantinuum's trapped ion devices \cite{illa2023multi-neutrino, amitrano2023trapped}, IBM's superconducting devices \cite{hall2021simulation, jhaa_2022, yeter-aydeniz_2022}, and D-Wave's {\tt Advantage} quantum annealer \cite{illa2022basic}. For a recent review of the entire field of collective neutrino oscillations, see Ref. \cite{patwardhan2023}.

The above literature has only studied neutrinos under the assumption that the number of possible neutrino flavors, $n_f$, is 2, with the physical $n_f$ = 3 collective neutrino oscillation picture being analyzed entirely in the classical approximation and in mean-field theory \cite{Fogli2009Supernova,Duan2008Flavor,Dasgupta2008Spectral,Dasgupta2009Multiple,Dasgupta2010Neutrino, Friedland2010SelfRefraction, sasaki2017possible, Airen2018, Chakraborty2020, Shalgar2021,Fiorillo2023a, martin2023equilibration} until 2023, when the first exact calculations of three-flavor collective neutrino oscillations were done \cite{siwach2023entanglement, balantekin2023quantum}. Ref. \cite{siwach2023entanglement} found a difference between results for $n_f$ = 3 and those for $n_f$ = 2.  
One of the limitations of the first two exact $n_f$ = 3 studies was that they used the single-angle approximation. Physically, neutrino-neutrino interactions depend on the angle between the trajectories of the interacting neutrinos. However, in the single-angle approximation, the trajectory-directions are averaged over \cite{duan2006simulation}. Supernova models that use the single-angle approximation exhibit several differences with those that do not \cite{wu2015effects, Duan2006, Esteban-Pretel2008, Wu2011, Raffelt2007}. For instance, the single-angle approximation produces collective neutrino oscillations at earlier times than the full (``multi-angle'') treatment of each neutrino's trajectory \cite{duan2011self}. These differences reflect themselves in supernova neutrino spectra \cite{sasaki2017possible, Mirizzi2011, Fogli2007, Banerjee2011, Dasgupta2009Multiple, Fogli2008, Fogli2007} and in nucleosynthesis \cite{duan2011influence}.  




Thus, it is important to the astrophysical applications of neutrino science to devise techniques for multi-angle simulations of collective neutrino oscillations with $n_f$ = 3, and to create algorithms by which future quantum computers can efficiently compute such simulations. This paper details the extension of the two-flavor collective neutrino oscillation analysis in Ref. \cite{illa2022basic} to three flavors.  The number of neutrinos in the simulation, $N$, is set to 2 for tests and  simple demonstrations and 4 for the main run.  While such system sizes are microscopic compared to what is possible with mean-field theory and are too small to have direct applications to astrophysical observables or supernova thermodynamics,  they can still provide insight into the entanglement structure evolution of dense neutrino systems and are comparable to the $N = 5$ choice in the exact $n_f$ = 3 simulations that have been done so far \cite{siwach2023entanglement, balantekin2023quantum}.  Furthermore, system sizes of this scale allow for testing of quantum algorithms that will enable the macroscopic simulation of full neutrino entanglement dynamics once quantum devices with sufficient qubit-count and state fidelity become available.

This paper is organized as follows. In Section \ref{sec:collneutoscham}, the Hamiltonian that defines the dynamics of the collective neutrino oscillations is established and discussed. In Section \ref{sec:dwaveimplementationdescription}, the techniques for implementing dynamics on D-Wave's {\tt Advantage} quantum annealer are described. Section \ref{sec:fourneutrinosim} presents the results of the simulation.  
Section \ref{sec:othersystems} discusses extension of the study to antineutrinos, Majorana neutrinos, and to time-dependent neutrino-neutrino interactions. Section \ref{sec:conclusion} discusses the implications of the results and future directions of research. Appendices discuss technical details.
 

\section{Collective Neutrino Oscillation Hamiltonian}
\label{sec:collneutoscham}

The terms relevant to the construction of a Hamiltonian describing the flavor dynamics of collective neutrino oscillations are the vacuum propagation (which effects neutrino oscillations), the neutrino-neutrino interactions \cite{fuller_resonant_1987, notzold_neutrino_1988, sigl_general_1993}, and the matter term (notably the Mikheyev-Smirnov-Wolfenstein (MSW) effect \cite{Wolfenstein1978, Mikheyev1985, Mikheev1986}). 
Following the leads of Refs. \cite{siwach2023entanglement, patwardhan2021spectral, cervia2022collective, lacroix2022role, pehlivan2011invariants, pehlivan2014neutrino, birol2018neutrino, cervia2019symmetries, patwardhan2019eigenvalues, rrapaj2020exact} the matter effects are assumed to be negligible.  Studies implementing a co-rotating frame have supported this approximation for the case of homogeneous systems modeled with the mean-field theory approach \cite{Duan2006b, Hannestad2006}.   The neutrino-neutrino terms are dependent on neutrino density \cite{qian1995neutrino} and hence are time-dependent in physical applications, as neutrinos radiate outward from a source and their density decreases as they do so. However, one of the primary aims of this study is to devise techniques for simulation on the {\tt Advantage} quantum annealer. For this reason, the Hamiltonian is treated as time-independent, following the lead of Ref. \cite{illa2022basic}. Additionally, only coherent interactions that either preserve or exchange the neutrino momenta (``coherent elastic forward scattering'') are taken into account. Discussion in the literature of interactions other than coherent elastic forward scattering has been taking place for decades \cite{pantaleone1992neutrino, pantaleone1992dirac, friedland2003manyparticle, Friedland2003Neutrino, Bell2003Speedup, Sawyer2004Instabilities, Friedland2006ManyBody, Balantekin2007, pehlivan2011invariants, birol2018neutrino, Cervia2019Entanglement, roggero2021entanglement,roggero2021dynamical,Martin2022Classical, xiong2022manybody, roggero2022entanglement, lacroix2022role, cervia2022collective, Martin2023ManyBody, Johns2023Neutrino, Fiorillo2024, cirigliano2024neutrino}. While the first quantum many-body study of collective neutrino oscillations that included the full forward and non-forward scattering was recently done \cite{cirigliano2024neutrino}, in this work the coherent forward scattering approximation is used, in the interest of a simple mapping of the problem onto quantum devices.

$n_f$ = 3 collective neutrino oscillations of N neutrinos in the mass basis can be represented in terms of the Gell-Mann matrices \cite{Balantekin2007, siwach2023entanglement}:

\begin{equation}
    H_{CNO} = \sum_{p = 1}^N \vec{B}(E_p) \cdot \vec{\lambda}_p + \sum_{p = 1}^{N - 1} \sum_{p' = p + 1}^{N} k (1 - \cos{\theta_{p p'}}) \vec{\lambda}_p \cdot \vec{\lambda}_{p^{'}}
    \label{eq:H_collectiveneutrinooscilation}
\end{equation}

\noindent where $\vec{\lambda}_p$ is the 8-term vector composed of the Gell-Mann matrices applied to the neutrino indexed by $p$, 
$\theta_{p p'}$ is the angle between the momenta of the neutrinos indexed by $p$ and $p'$, $k$ is the coupling strength of the neutrino-neutrino interaction, $E_p$ is the energy of the neutrino indexed by p, and $\vec{B} (E_p)$ is a vector that is a function of $E_p$ and that creates the diagonalized single-neutrino oscillation term when dotted with $\vec{\lambda}_p$. The diagonalized single-vector neutrino oscillation term is determined by two parameters: $\delta m^2$, the difference in mass-squared between the first and second mass eigenstates, and $ \Delta m^2$, the difference between the third mass eigenstate's mass-squared and the mean of the masses-squared of the first and second mass eigenstates (following the convention of Ref. \cite{capozzi2014status}).  Normal mass hierarchy is assumed throughout this work. It turns out that two neutrinos with the same starting state and with the same momentum behave exactly the same as if they were just one neutrino. Hence, following the lead of Ref. \cite{siwach2023entanglement}, a neutrino referenced by p can be thought of as a momentum-mode. In order to convert between the flavor and mass basis, the Pontecorvo–Maki–Nakagawa–Sakata (PMNS) matrix \cite{Maki1962, Pontecorvo1957, Schechter1980} is utilized:

\begin{equation}
\label{eq:PMNS3flav_definition}
U_{PMNS} = 
\begin{pmatrix}
1 & 0 & 0 \\
0 & c_{23} & s_{23} \\
0 & -s_{23} & _{23} \\
\end{pmatrix}
\begin{pmatrix}
c_{13} & 0 & s_{13}e^{-i \delta} \\
0 & 1 & 0 \\
-s_{13}e^{i \delta} & 0 & c_{13} \\
\end{pmatrix}
\begin{pmatrix}
c_{12} & s_{12} & 0 \\
-s_{12} & c_{12} & 0 \\
0 & 0 & 1 \\
\end{pmatrix}
\end{equation}

\noindent where $c_{ij} = \cos{\theta_{ij}}$ and $s_{ij} = \sin{\theta_{ij}}$. One feature of Eq. \ref{eq:H_collectiveneutrinooscilation} is that as long as the absolute values of the neutrino energies are all the same, the single-neutrino term commutes with the neutrino-neutrino interaction term. Related to this, the neutrino-neutrino interaction term is independent of the basis that the neutrinos are in, so long as the neutrinos are all in the same basis. This fact enables the strategy utilized in Sec. \ref{sec:fourneutrinosim} to work and has several consequences for the observations therein. 


The parameters used in this project can be found in Tab. \ref{tab:2flavexactsimulationpars}. The single-neutrino term parameters are drawn from Ref. \cite{siwach2023entanglement} and fall within the 1 $\sigma$ confidence interval of experimental results if rounded to three significant figures as of the release of the 2023 PDG Review \cite{workman2023particle}. Generally, the single-neutrino oscillation term vector $\vec{B}(E_p)$ is defined in terms of the parameters in Tab. \ref{tab:2flavexactsimulationpars}. In this work, unless it is either stated otherwise or the choice of $\vec{B}(E_p)$ is arbitrary, $\vec{B}(E_p)$ is set to its physical value, as derived in App. \ref{app:singlenuosc}: 

\begin{equation}
    \vec{B}(E_p) \rightarrow (0, 0, -\frac{\delta m^2}{4 E_p}, 0, 0, 0, 0, -\frac{\Delta m^2}{2\sqrt{3} E_p})
    \label{eq:BEpdefinition}
\end{equation}

\noindent For N = 4, the angles $\theta_{ij}$ between the neutrinos, used for the neutrino-neutrino interaction term, are given by the anisotropic angle distribution from Ref. \cite{illa2022basic}, with $\xi$ = 0.9:

\begin{equation}
    \label{eq:anisotropicangles}
    \theta_{ij} = \arccos{(\xi)} \frac{\abs{i - j}}{N - 1}
\end{equation}



\begin{table}[ht]
\centering

\begin{tabular}{c|c|c}
\hline
Parameter & Values & PDG Experimental results \\
\hline
$E_i$ & $10^7$ $eV$ & \\
$\delta m^2$ & $7.42  \times 10^{-5}$ $eV^2$ & $(7.53(18)) \times 10^{-5}$ $eV^2$ \\
$\Delta m^2$ & $2.44  \times 10^{-3}$ $eV^2$ & $(2.475(33)) \times 10^{-3}$ $eV^2$\\
${\theta_v}_{12}$ & 0.591667 & $0.587(14)$\\
${\theta_v}_{13}$ & 0.148702 & $0.1489(24)$\\
${\theta_v}_{23}$ & 0.840027 & $0.832(^{+18}_{-12})$\\

$\delta$ & 4.36681 & $3.86(66)$ \\
$\theta_{12}$ (N = 2) & $\frac{\pi}{4}$ &  \\
$\theta_{ij}$ (N = 4) & Eq. \ref{eq:anisotropicangles} &  \\
$k$ & $1.75 \times 10^{-12}$ $eV$ &  \\ 

\end{tabular}
\caption{The parameters used in the exact simulation of $n_f$ = 3 interacting neutrino systems described by Eq. \ref{eq:H_collectiveneutrinooscilation} as well as the experimental results from the 2023 PDG Review \cite{workman2023particle} for neutrino oscillation terms. For all $n_f$ = 2 studies, unless otherwise stated, the mass-difference is set to $\Delta m^2$ and the mixing angle to ${\theta_v}_{12}$. Normal mass hierarchy is assumed.}

\label{tab:2flavexactsimulationpars}

\end{table}



\section{Implementation on D-Wave {\tt Advantage} Quantum Annealer}
\label{sec:dwaveimplementationdescription}

D-Wave's {\tt Advantage} quantum annealer is a 5000+-qubit device \cite{mcgeoch2022advantage} designed with the express function of obtaining the ground state of a user-specified classical Ising model through a procedure known as quantum annealing. In quantum annealing, the system begins in the ground state of one Hamiltonian $H_i$ and is time-evolved on a Hamiltonian $H_a$ that begins as $H_i$ but gradually becomes a different Hamiltonian, $H_f$. As long as $H_a$'s transition from $H_i$ to $H_f$ is sufficiently slow, the system's final state will be the ground state of $H_f$ \cite{kadowaki1998quantum}.
For {\tt Advantage}, $H_i = (\sum_i \hat{\sigma}^{(i)}_x)$ and $H_f = (\sum_i h_i \hat{\sigma}^{(i)}_z + \sum_{i > j} J_{ij} \hat{\sigma}^{(i)}_z \hat{\sigma}^{(j)}_z )$, where $\hat{\sigma}^{(i)}_x$ and $\hat{\sigma}^{(i)}_z$ are the Pauli-x and Pauli-z matrices, respectively, acting on the $i^{th}$ qubit, and $h_i$ and $J_{ij}$ are the user-tunable parameters. In turn, $H_a$ is set to the following Hamiltonian \cite{dwavesysdocs}:

\begin{equation}
    H_{ising} = \frac{A(s)}{2} (\sum_i \hat{\sigma}^{(i)}_x) + \frac{B(s)}{2}(\sum_i h_i \hat{\sigma}^{(i)}_z + \sum_{i > j} J_{ij} \hat{\sigma}^{(i)}_z \hat{\sigma}^{(j)}_z )
    \label{eq:isingmodeltunable}
\end{equation}

\noindent where A(s) and B(s) are time-dependent parameters. s is a time-parameter, defined as $s = t/t_f$, where t is time and $t_f$ is the total time of the anneal. In order for quantum annealing to be executed, A(s) and B(s) are defined so that $A(0) >> B(0)$ and $A(1) << B(1)$. The exact definitions of A(s) and B(s) are known as the annealing schedule. After the anneal is done, all of the qubits on {\tt Advantage} are measured in the Pauli-Z basis to obtain a result for $H_f$'s ground state. This result is returned alongside its expectation value of $H_f$ \cite{dwavesysdocs}.


One of the main documentation-prescribed methods of mapping an optimization problem onto {\tt Advantage} is to map it onto a QUBO (quadratic unconstrained binary optimization) problem \cite{Lucas2014, dwavesysdocs, mcgeoch2022advantage}. A QUBO problem is one that involves the minimization of a function of the form
\begin{equation}
    f(q) = \sum_i Q_{ii} q_i + \sum_{i < j} Q_{ij} q_i q_j
    \label{eq:QUBOproblemdefinition}
\end{equation}

\noindent with $q_i$ being a binary variable with 2 possible values: 0 or 1.

QUBO problems are directly mappable onto {\tt Advantage}'s $H_f$ \cite{mcgeoch2022advantage} and one can directly submit a problem with a given set of $Q_{ij}$ values to the Ocean interface provided by DWave Systems, which will convert it to a set of $h_i$ and $J_{ij}$ parameters to be submitted directly to the annealer. The annealing process is then done, and the measurement-result of each qubit on {\tt Advantage} is treated as a result for the corresponding binary variable on the QUBO problem \cite{dwavesysdocs}. 
 
\subsection{Mapping time evolution onto a QUBO problem}
\label{sec:timeevontoQUBOmapping}
 {\tt Advantage} is an optimizer, so to conduct a time-evolution simulation on it, the time-evolution must first be mapped to an optimization problem. In this work, this mapping is accomplished using the Feynman clock Hamiltonian (``C")\cite{mcclean2013feynman}:
\begin{equation}
    C = C_0 + \frac{1}{2}\sum_t (I \otimes \ket{t} \bra{t} - U_t \otimes \ket{t+dt} \bra{t} - U_t^\dagger \otimes \ket{t} \bra{t+dt} + I \otimes \ket{t+dt} \bra{t+dt})
    \label{eq:feynmanclockham}
\end{equation}

\noindent In the Feynman clock Hamiltonian approach,  the time-evolution is discretized into time-steps separated by increment $dt$.  One register of states is assigned to each time-step and represents the system at that time-step. The Hamiltonian itself is designed so that the penalty term, $C_0$,  ensures that the desired initial state $\ket{\Psi_0}$ is the ground state of the initial time register and the other terms ensure that the ground state of the each of the other registers is equivalent to the result of applying the time-evolution operator,  $U_t = e^{-i H dt}$,  to the state on the previous time-step's register.

Thus, the desired result of the computations on {\tt Advantage} in this project is the ground state of the Feynman clock Hamiltonian that describes the time evolution of the system in consideration. The algorithm used to obtain these ground states is the adaptive quantum annealing eigensolver (AQAE), used previously in Refs. \cite{Chang2019, Rahman2021, illa2022basic}. In AQAE, one starts with a prior on the Feynman clock Hamiltonian's ground state, and the Feynman clock Hamiltonian is mapped to a QUBO problem which is sent to an annealer, and the annealer's measurement results are used to update the prior on the ground state. This process is done repeatedly, and at each repetition the magnitude of the update to the prior is reduced so that later repetitions improve the precision of the eventual result. In this project, AQAE is used to generate time-evolution under a Hamiltonian $H$ from an initial state $\ket{\Psi_0}$ at initial time $t_0$ to a final state $\ket{\Psi_f}$ at time $t_f$ using the following process:

\begin{enumerate}
    \item Map $H$ to a two-time-step Feynman clock Hamiltonian $H^{F}$ using Eqn. \ref{eq:feynmanclockham} with $C_0 = 0$. The penalty term will be added later, in the form of the third term in the mapping to the QUBO problem in Steps 4 and 7, which is equal to $Q_0$ from Eq. \ref{eq:QUBOpenaltyterm_app} from App. \ref{app:AQAEderivdemonstration}.
    \item Set the prior $a^{(0)}$ for the ground state of the Feynman clock Hamiltonian equal to $\ket{\Psi_0} \otimes \ket{t_i} + \ket{0} \otimes \ket{t_f}$, where $\ket{0}$ is the state with the same vector-space size as $\ket{\Psi_0}$ with a norm of 0.
    \item Define the variable $z$ and set it to 0.
    \item Map the Feynman clock Hamiltonian and the prior on the ground state to a QUBO $Q^{F1}_{AQAE}$ problem with the following formula:
    \begin{multline}
    Q^{F1}_{AQAE} = 2^{-1+\delta_{z0}} \left( \begin{array}{ccc|ccc}
    Re(H^{F}_{11}) & \ldots & Re(H^{F}_{1n}) & -Im(H^{F}_{11}) & \ldots & -Im(H^{F}_{1n}) \\
    \vdots & \ddots & \vdots & \vdots & \ddots & \vdots \\
    Re(H^{F}_{n1}) & \ldots  & Re(H^{F}_{nn}) & -Im(H^{F}_{n1}) & \ldots  & -Im(H^{F}_{nn}) \\
    \hline 
    Im(H^{F}_{11}) & \ldots & Im(H^{F}_{1n}) & Re(H^{F}_{11}) & \ldots & Re(H^{F}_{1n}) \\
    \vdots & \ddots & \vdots & \vdots & \ddots & \vdots \\
    Im(H^{F}_{n1}) & \ldots  & Im(H^{F}_{nn}) & Re(H^{F}_{n1}) & \ldots  & Re(H^{F}_{nn}) \\
\end{array} \right) \otimes \\ \begin{pmatrix}
    2^{-K + 1} 2^{-K + 1} & \ldots & 2^{-K + 1} 2^{-1} & - 2^{-K + 1} 2^{0} \\
    \vdots & \ddots & \vdots & \vdots \\
    2^{-1} 2^{-K + 1} & \ldots & 2^{-1} 2^{-1} & - 2^{0} 2^{0} \\
    -2^{0} 2^{-K + 1} & \ldots & -2^{0} 2^{-1} &  2^{0} 2^{0} 
    \end{pmatrix} + \left( \begin{array}{c|c}
    diag(Re(a^{(z)^*} H^{F})) & 0_{n,n} \\
    \hline 
    0_{n,n} & diag(-Im(a^{(z)^*} H^{F})) \\
\end{array} \right) \otimes \\
diag(2^{-K + 2 + z}, 2^{-K + 3 + z}, \ldots , 2^{z} , -2^{z + 1}) + diag(1, 0, 1, 0) \otimes 3.5 \mathbb{I}_{Kn, Kn}
    \label{eq:QFeynman_AQAEforwardorder}
\end{multline}   
Here, $Re()$ and $Im()$ refer to real and imaginary components, respectively, $*$ refers to complex conjugation, $K$ is an optimization parameter that refers to the number of binary variables in the QUBO problem submitted to {\tt Advantage} per dimension of the statevector of $H^F$'s ground state, and $n$ is the number of rows in $H^{F}$.
    \item Submit $Q^{F1}_{AQAE}$ to {\tt Advantage} for annealing, and obtain a string of qubit-measurements of the form $(q^1_1, \ldots, q^1_K, q^2_1, \ldots, q^2_K, \ldots, q^n_1, \ldots, q^n_K)$, which are results for values of $Q^{F1}_{AQAE}$'s QUBO problem's binary variables.
    \item Use the solution from Step 5 and the prior $a^{(z)}$ to obtain a posterior result $b^{(z)}$ of $H^{F}$'s ground state, like so:
    \begin{equation}
        b^{(z)}_\alpha = a^{(z)}_\alpha + (- 2^{-z} q^{\alpha}_K + \sum_{i = 1}^{K - 1} \frac{q^{\alpha}_i}{2^{K - i -z}}) + i(- 2^{-z} q^{\alpha + n}_K + \sum_{i = 1}^{K - 1} \frac{q^{\alpha + n}_i}{2^{K - i -z}})
        \label{eq:azposteriorfindingnormaldig}
    \end{equation}
    \item Map the Feynman clock Hamiltonian and the prior on the ground state to a QUBO $Q^{F2}_{AQAE}$ problem with the following formula:
    \begin{multline}
    Q^{F2}_{AQAE} = 2^{-1+\delta_{z0}} \left( \begin{array}{ccc|ccc}
    Re(H^{F}_{11}) & \ldots & Re(H^{F}_{1n}) & -Im(H^{F}_{11}) & \ldots & -Im(H^{F}_{1n}) \\
    \vdots & \ddots & \vdots & \vdots & \ddots & \vdots \\
    Re(H^{F}_{n1}) & \ldots  & Re(H^{F}_{nn}) & -Im(H^{F}_{n1}) & \ldots  & -Im(H^{F}_{nn}) \\
    \hline 
    Im(H^{F}_{11}) & \ldots & Im(H^{F}_{1n}) & Re(H^{F}_{11}) & \ldots & Re(H^{F}_{1n}) \\
    \vdots & \ddots & \vdots & \vdots & \ddots & \vdots \\
    Im(H^{F}_{n1}) & \ldots  & Im(H^{F}_{nn}) & Re(H^{F}_{n1}) & \ldots  & Re(H^{F}_{nn}) \\
\end{array} \right) \otimes \\ \begin{pmatrix}
    2^{-K + 1} 2^{-K + 1} & \ldots & 2^{-K + 1} 2^{-1} & - 2^{-K + 1} 2^{0} \\
    \vdots & \ddots & \vdots & \vdots \\
    2^{-1} 2^{-K + 1} & \ldots & 2^{-1} 2^{-1} & - 2^{0} 2^{0} \\
    -2^{0} 2^{-K + 1} & \ldots & -2^{0} 2^{-1} &  2^{0} 2^{0} 
    \end{pmatrix} - \left( \begin{array}{c|c}
    diag(Re(b^{(z)^*} H^{F})) & 0_{n,n} \\
    \hline 
    0_{n,n} & diag(-Im(b^{(z)^*} H^{F})) \\
\end{array} \right) \otimes \\
diag(2^{-K + 2 + z}, 2^{-K + 3 + z}, \ldots , 2^{z} , -2^{z + 1}) + diag(1, 0, 1, 0) \otimes 3.5 \mathbb{I}_{Kn, Kn}
    \label{eq:QFeynman_AQAEreverseorder}
\end{multline}
    \item Submit $Q^{F2}_{AQAE}$ to {\tt Advantage} for annealing, and obtain a string of qubit-measurements of the form $(q^1_1, \ldots, q^1_K, q^2_1, \ldots, q^2_K, \ldots, q^n_1, \ldots, q^n_K)$, which are results for values of $Q^{F2}_{AQAE}$'s QUBO problem's binary variables.
    \item Use the solution from Step 8 and the prior $b^{(z)}$ to obtain a posterior result $a^{(z+1)}$ of $H_f$'s ground state, like so:
    \begin{equation}
        a^{(z + 1)}_\alpha = b^{(z)}_\alpha + (2^{-z} q^{\alpha}_K - \sum_{i = 1}^{K - 1} \frac{q^{\alpha}_i}{2^{K - i -z}} ) + i(2^{-z} q^{\alpha + n}_K - \sum_{i = 1}^{K - 1} \frac{q^{\alpha + n}_i}{2^{K - i -z}} ) 
        \label{eq:azposteriorfindingreversedig}
    \end{equation}
    \item Increment $z$ by 1.
    \item Repeat Steps 4-10 until the result $a^{(z)}$ converges. After convergence, $a^{(z)}$ should equal $\ket{\Psi_0} \otimes \ket{t_0} + \ket{\Psi_g} \otimes \ket{t_f}$
\end{enumerate}



One limitation of AQAE is that it cannot obtain quantum advantage for time evolution.
This is because its construction of the QUBO matrix submitted to the device at each iteration involves the matrix-multiplication of the state-vector obtained from the previous iteration, which is a task of equal computational difficulty to classically obtaining the time-evolution. However, without AQAE's iterative approach, obtaining results from {\tt Advantage} to precision $\epsilon$ requires the use of $O(\lceil \log_2(\frac{1}{\epsilon}) \rceil)$ qubits per real number in the results. AQAE's iterative process, on the other hand, only requires 1-2 qubits per real number for all levels of precision and thus allows state-of-the-art quantum annealers to obtain results to double-type precision for problems that they would otherwise not have had the qubits needed to solve to a precision greater than one significant figure. Additionally, because later iterations can correct for bit-flip errors in previous iterations, AQAE is resilient to noise, which is invaluable on state-of-the art quantum annealers, such as {\tt Advantage}, which have a large amount of noise. Thus, AQAE is currently necessary for development of quantum annealing algorithms for high-complexity problems.  A detailed derivation of AQAE from more basic quantum annealing methods with a demonstration using a simple one-neutrino system can be found in App. \ref{app:AQAEderivdemonstration}. In App. \ref{app:nealbenchmarking}, a full test of this process on both {\tt neal}, DWave, Inc.'s provided classical thermal annealer, and on the {\tt Advantage} annealer for a two-neutrino system is discussed. 


\section{N = 4 time evolution simulations}
\label{sec:fourneutrinosim}
A system's neutrino-count N must be at least 3 in order to capture the physical differences between neutrino-neutrino interactions in the case where the number of flavors, $n_f$, is 2 and those in the case where $n_f$ = 3. However, time-evolution of systems with 30 basis-states failed to converge on {\tt Advantage}. 20 basis-state systems mostly annealed like normal but did have a noticeably higher frequency stuck at a local minimum before convergence. This presents a challenge, because the Hilbert-space size of a system with N = 3 and $n_f$ = 3 is 27, which is within the range at which the time-evolution fails to reach convergence on the annealer. 


To circumvent this, domain-decomposition is used. This technique, which has been used in applications of the quantum annealer to finding the ground state of non-Abelian quantum field theories \cite{Rahman2021, Farrell2023a}, takes advantage of the block-diagonalized structure of Hamiltonians to find the solution for each block separately. The mass-basis collective neutrino oscillation Hamiltionian is natively block-diagonalized: it doesn't change the number of neutrinos in each mass-eigenstate. An N = 5, $n_f$ = 3 system could be encoded onto {\tt Advantage} in this way, but could not be simulated because the largest subsystem produced in this case would be a 30-state space.


Hence, this technique was used to find the time-evolution of the quantum entanglement of an $n_f$ = 3, N = 4 system with an initial state of $\ket{\nu_{e} \nu_{e} \nu_{\tau} \nu_{\mu}}$. The motivation for finding the entanglement is that entanglement is the quantity which indicates intractability with classical devices \cite{vidal2003efficient, VanDenNest2006, Yoran2007, VanDenNest2007, VanDenNest2007NJP}. This system is analogous to the $\ket{\nu_{e} \nu_{e} \nu_{\mu} \nu_{\mu}}$ initial state used for the $n_f$ = 2, N = 4 simulation in Ref. \cite{illa2022basic}. The third neutrino in the system, which for an initial state of $\ket{\nu_{e} \nu_{e} \nu_{\mu} \nu_{\mu}}$ is the neutrino starting out in the $\nu_\tau$ state, is treated as a probe interacting with a system of $\nu_e$ and $\nu_\mu$. Since (as discussed in Sec. \ref{sec:collneutoscham}), Eq. \ref{eq:H_collectiveneutrinooscilation} commutes for systems where all neutrinos have the same energy magnitude, it seems intuitive that the single-neutrino oscillation term has no effect on the entanglement witnesses. This is because due to the commutation, all of the time-evolution from the single-neutrino oscillation can be pushed to the end of the time-evolution, where it would have no effect on the entanglement, and ostensibly, the entanglement witnesses as well. Thus, entanglement only depends on the neutrino-neutrino interaction term, as long as all the energies of the individual neutrinos are the same.

It is possible to conduct this simulation with neutrinos of different energies using the same methods at no extra resource cost. One caveat is that the single-neutrino oscillation interaction will likely have an effect on the entanglement, due to the fact that in this case the single-neutrino interaction does not commute with the neutrino-neutrino interaction in this case.  However, in this study the neutrino energies $E_p$ are fixed to the same value (denoted as E from this point forward) in order to compare to the results in Ref. \cite{illa2022basic}. The Hamiltonian parameters can be found in Tab. \ref{tab:2flavexactsimulationpars}. 

The first step was the simulation of the system using exact classical numerical methods and comparison between results for the case where the physical-case parameters from Tab. \ref{tab:2flavexactsimulationpars} are used and results for the cases where these parameters are sequentially set to 0, as well as to analogous simulations for initial states of $\ket{\nu_{e} \nu_{e} \nu_{\mu} \nu_{\mu}}$ and $\ket{\nu_{e} \nu_{e} \nu_{e} \nu_{\mu}}$. Additionally, the classical simulations for $\ket{\nu_{e} \nu_{e} \nu_{\mu} \nu_{\mu}}$ were repeated with a Hamiltonian for $n_f = 2$  that in the mass-basis works out to \cite{illa2022basic}


\begin{equation}
    H_{CNO} (n_f = 2) = \sum_{p = 1}^N \begin{pmatrix} -\frac{\Delta m^2}{4 E} & 0 \\ 0 & \frac{\Delta m^2}{4 E} \end{pmatrix}_p + \sum_{p = 1}^{N - 1} \sum_{p' = p + 1}^{N} k (1 - \cos{\theta_{p p'}}) \vec{\sigma}_p \cdot \vec{\sigma}_{p^{'}}
    \label{eq:H_collectiveneutrinooscilation_nf2}
\end{equation}

\noindent where $\vec{\sigma}_p = (\hat{\sigma}^x_p, \hat{\sigma}^y_p, \hat{\sigma}^z_p)$; $\hat{\sigma}^x_p$, $\hat{\sigma}^y_p$, and $\hat{\sigma}^z_p$ are the Pauli-x, Pauli-y, and Pauli-z operators, respectively, acting on the $p^{th}$ neutrino; and $\Delta m^2$ is the larger squared mass difference from Tab. \ref{tab:2flavexactsimulationpars}. The subscript on the 2x2 matrix (which is the single neutrino oscillation term) denotes the index of the neutrino on which it acts on. I attempted substituting $\Delta m^2$ for $\delta m^2$ from Tab. \ref{tab:2flavexactsimulationpars} and applying three different basis-changes to $H_{CNO}(n_f = 2)$. Each basis change was done by applying the same SO(2) transformation to each neutrino in the system. The SO(2) rotation angle of each basis-change was a different one of the mixing angles in Tab. \ref{tab:2flavexactsimulationpars}. These SO(2) transformations are the $n_f = 2$ equivalent to the PMNS matrix \cite{Maki1962, Pontecorvo1957, Schechter1980}. 

The entanglement entropy of each neutrino and the negativity between each pair of neutrinos was then extracted for each sample. The entanglement entropy, which is a measure of entaglement between a given neutrino (labeled with index $i$) and all other neutrinos, is calculated as a von Neumann entropy \cite{schumacher1995quantum}:

\begin{equation}
    S_i(t) = - Tr(\rho_i \log_2(\rho_i))
    \label{eq:entanglemententropydefinition}
\end{equation}

\noindent where $\rho_i$ is the one-neutrino reduced density matrix for the $i^{th}$ neutrino. The logarithmic negativity, which is a measure of entanglement between two specific neutrinos (labeled with indices $i$ and $j$ in this case) is calculated like so \cite{zyczkowski1998volume, vidal2002computable}:

\begin{equation}
    N_{ij}(t) = \log_2||\rho^{\Gamma}_{ij}(t)||_1
    \label{eq:negativitycalc}
\end{equation}

\noindent where $\rho_{ij}$ is the reduced density matrix for the $ij$ neutrino pair, $\Gamma$ indicates partial transposition of the matrix it is a superscript of, and $|| \cdot ||_1$ is the trace norm.

\begin{figure}[ht]
    \raggedleft
\includegraphics[width=0.05\textwidth]{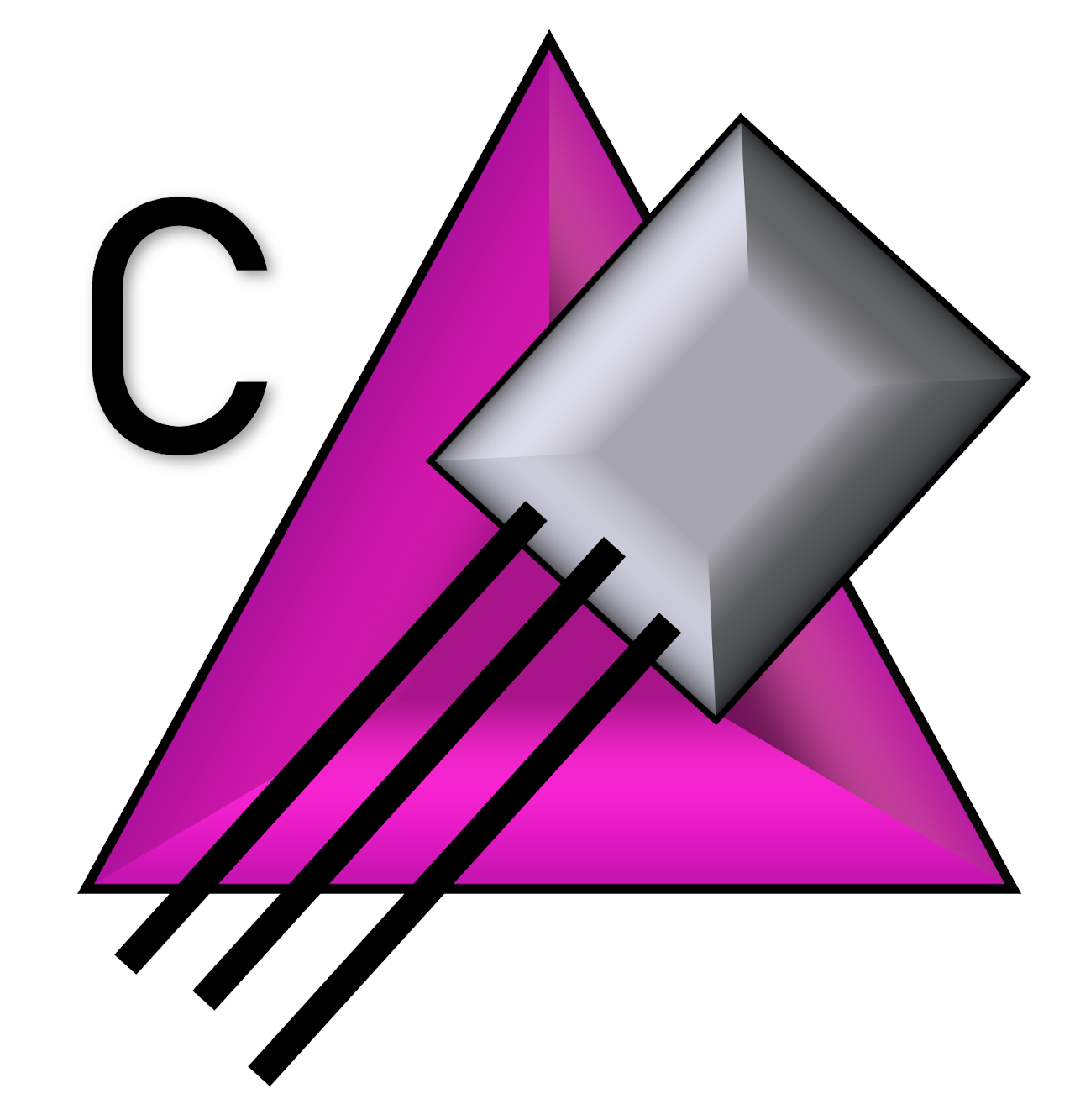}

    \centering
    \includegraphics[width=0.20\textwidth]{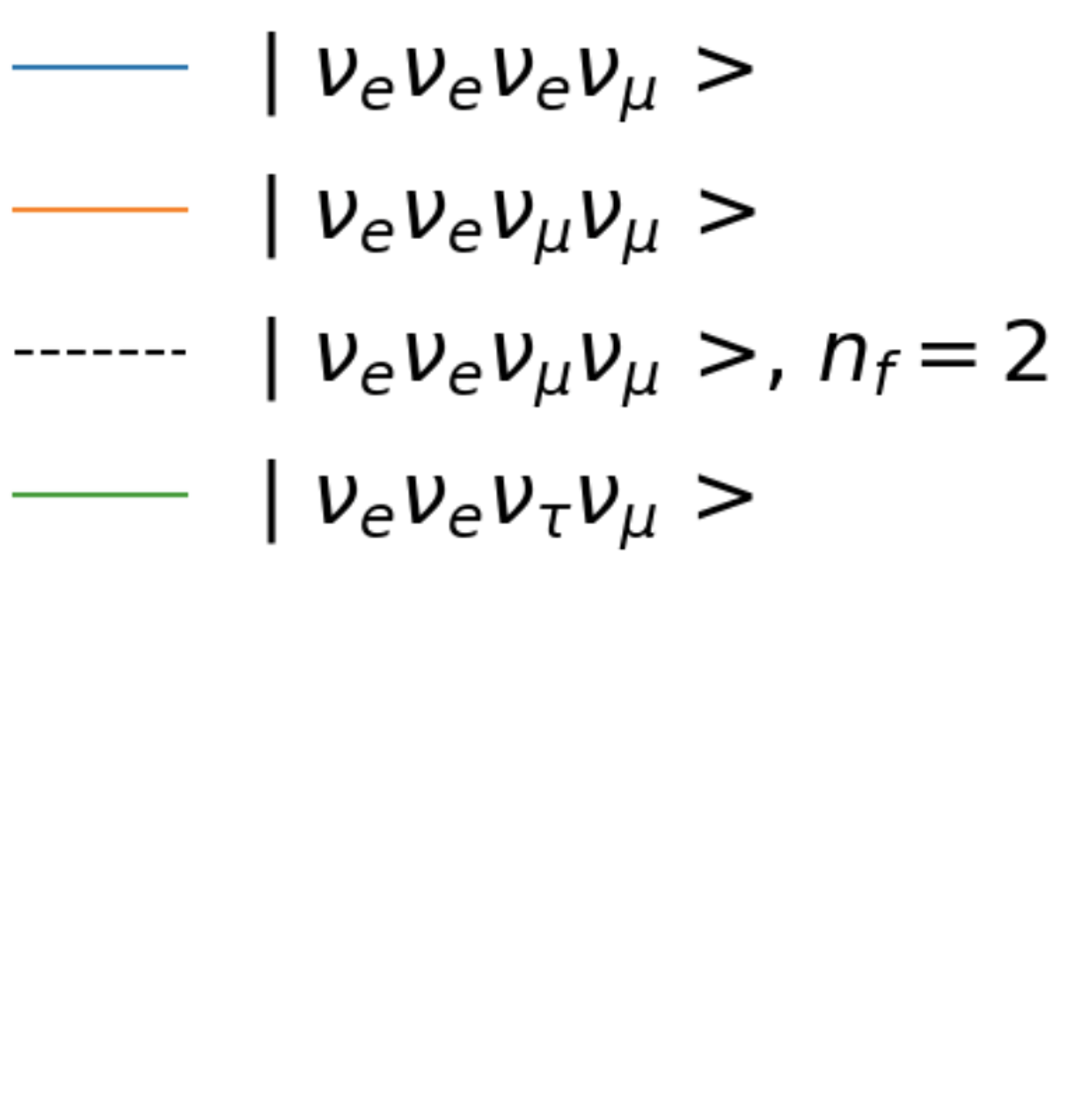}
    \includegraphics[width=0.32\textwidth]{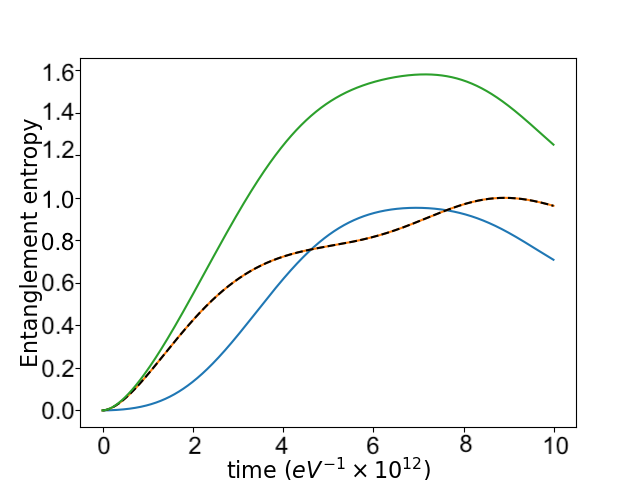} \\
    \includegraphics[width=0.32\textwidth]{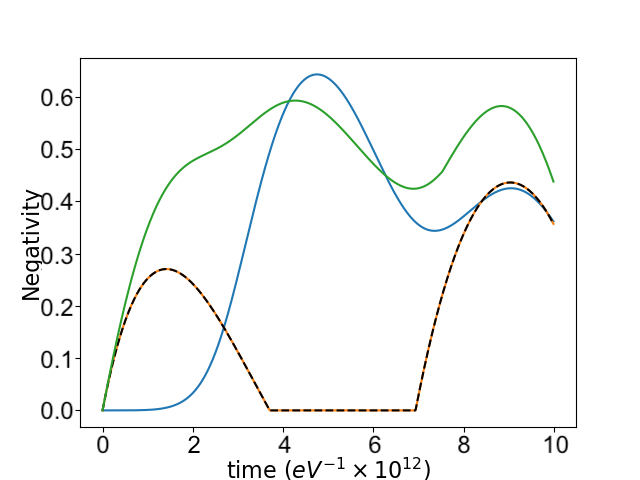}
    \includegraphics[width=0.32\textwidth]{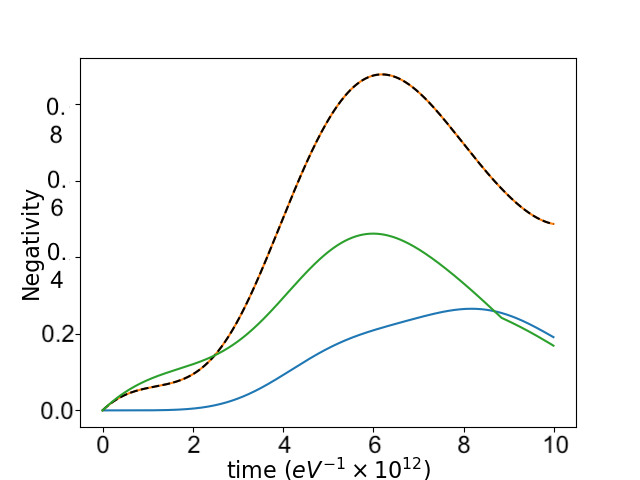}
    \includegraphics[width=0.32\textwidth]{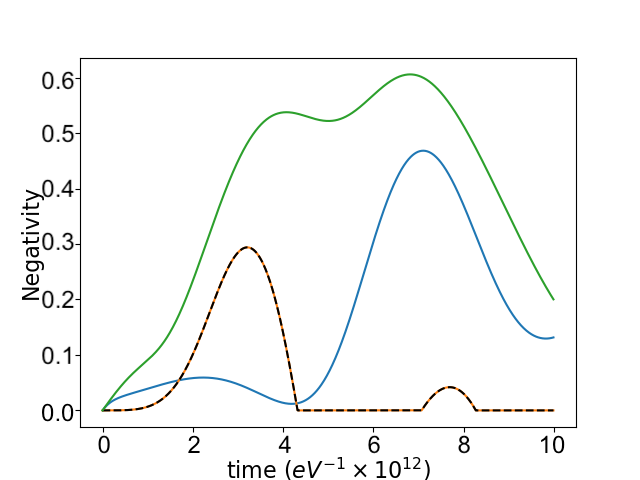} \\
    \caption{Under collective neutrino oscillations, time evolution of: entanglement entropy of the third neutrino (top), negativity between neutrinos 1 and 3 (bottom left), negativity between neutrinos 2 and 3 (bottom middle), and negativity between neutrinos 3 and 4 (bottom right). The terms in the $\nu$-$\nu$ interaction term of the Hamiltonian are given by Tab. \ref{tab:2flavexactsimulationpars}. Number of neutrinos in the simulation, N, is 4 and number of neutrino-flavors, $n_f$, is 3. The purple icon at the top-right indicates exact simulation on a classical device \cite{Klco2020}.}
    \label{fig:classical_probenuresults_Nequals4}
\end{figure}

An additional part of this step is the verification of the expectation that the evolution of entanglement witnesses in the case that the neutrino energies are all the same is dependent only on the neutrino-neutrino interaction term. I did this by repeating this study for the following four choices of $\vec{B}(E)$ from the $n_f = 3$ collective neutrino oscillation Hamiltonian from Eq. \ref{eq:H_collectiveneutrinooscilation}: 

\begin{itemize}
    \item $\vec{B}(E) = (0, 0, 0, 0, 0, 0, 0, 0)$, i.e. no single-neutrino flavor oscillation
    \item $\vec{B}(E) = (0, 0, -\frac{\delta m^2}{4 E}, 0, 0, 0, 0, -\frac{\Delta m^2}{2\sqrt{3} E})$, from the derivation of the ultrarelativistic-limit time-evolution of a generic three-mass-eigenstate system in App. \ref{app:singlenuosc}
    \item $\vec{B}(E) = (0, 0, -\frac{\delta m^2}{12 E}, 0, 0, 0, 0, -\frac{\Delta m^2}{6\sqrt{3} E})$
    \item $\vec{B}(E) = (0, 0, -\frac{\delta m^2}{4 E}, 0, 0, 0, 0, -\frac{\Delta m^2}{4 E})$, from Ref. \cite{patwardhan2023} 
    
\end{itemize}

\noindent all four choices of $\vec{B}(E)$ produced the results in Fig. \ref{fig:classical_probenuresults_Nequals4}. For $n_f = 2$, I found the same apparent invariance of results for entanglement entropy and negativity with the single-neutrino oscillation term that I found for $n_f = 3$ by comparing the time-evolution results for $H_{CNO} (n_f = 2)$ as expressed in Eq. \ref{eq:H_collectiveneutrinooscilation_nf2} to results where the $\Delta m^2$ in $H_{CNO} (n_f = 2)$ was replaced with $\delta m^2$ from Tab. \ref{tab:2flavexactsimulationpars} and to results where $\Delta m^2$ was replaced with 0. This supports prediction that only the neutrino-neutrino interaction term affects the values of the entanglement witnesses.

Additionally the results for an initial state of $\ket{\nu_{e} \nu_{e} \nu_{\mu} \nu_{\mu}}$ were the same for $n_f = 2$ and $n_f = 3$. Hence, the only differences found were based on the initial state, specifically where there are 3 flavors in the initial state. The results for the probe-neutrino's entanglement entropy and negativities with each other neutrino are shown in Fig. \ref{fig:classical_probenuresults_Nequals4}. Generally, the results for each initial state were distinct, but there were a few highlights. First, the probe's entanglement entropy exhibits a similar peak for an initial state of $\ket{\nu_{e} \nu_{e} \nu_{\tau} \nu_{\mu}}$ as for an initial state of $\ket{\nu_{e} \nu_{e} \nu_{e} \nu_{\mu}}$, though at a higher magnitude, reflecting the fact that the maximum magnitude of entanglement entropy is $log_2 (3)$ for $n_f$ = 3 and 1 for $n_f$ = 2. The negativities for the $\ket{\nu_{e} \nu_{e} \nu_{\tau} \nu_{\mu}}$ seemed to exhibit the peaks of both the negativities of $\ket{\nu_{e} \nu_{e} \nu_{\mu} \nu_{\mu}}$ and the negativities of $\ket{\nu_{e} \nu_{e} \nu_{e} \nu_{\mu}}$. Although this result is pretty abstract, it nonetheless suggests a degree of unique behavior from the introduction of a third flavor into an interacting neutrino system. 

\begin{figure}
    \raggedleft
    \includegraphics[width=0.05\textwidth]{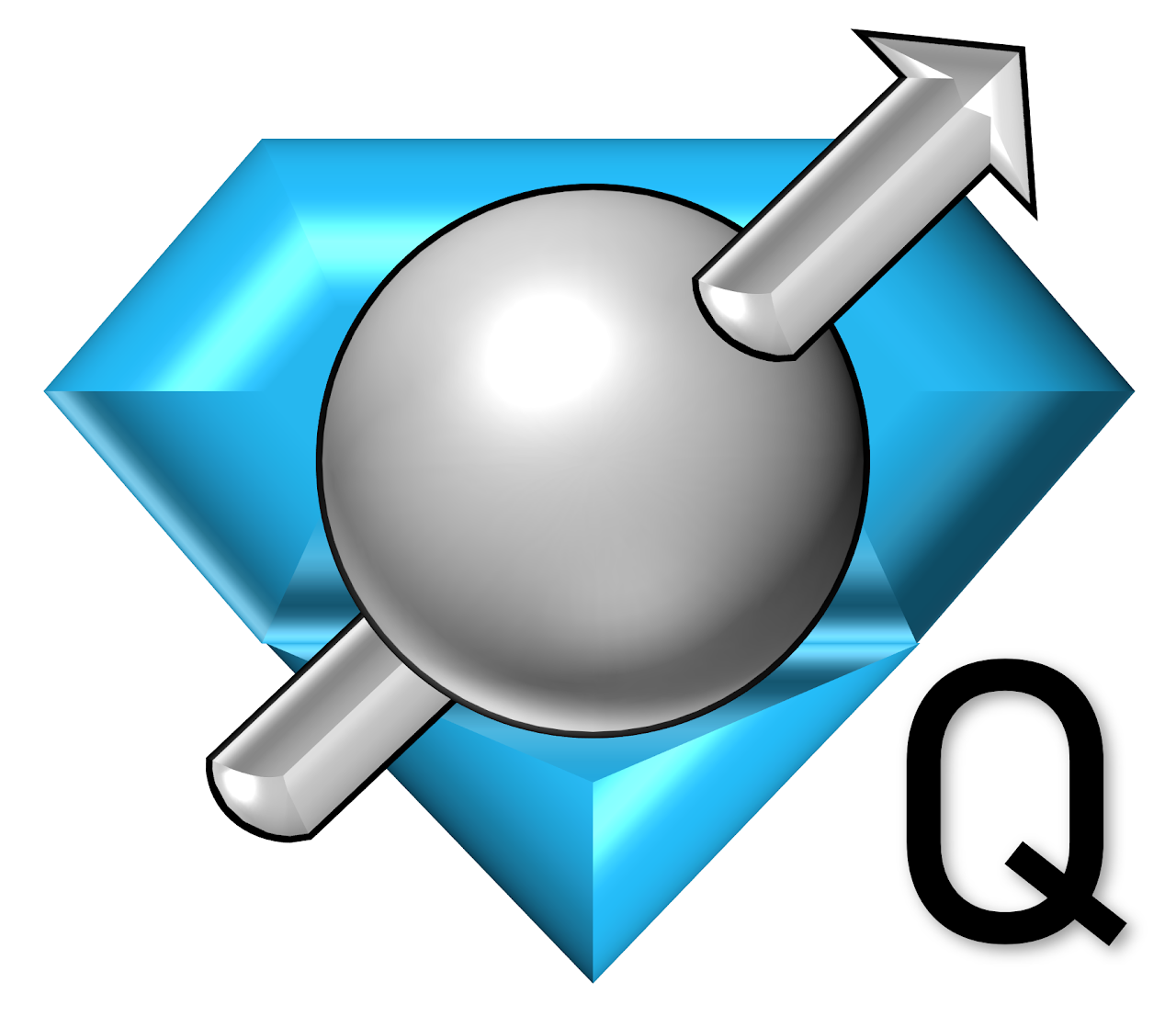}

    \centering
    \includegraphics[width=0.20\textwidth]{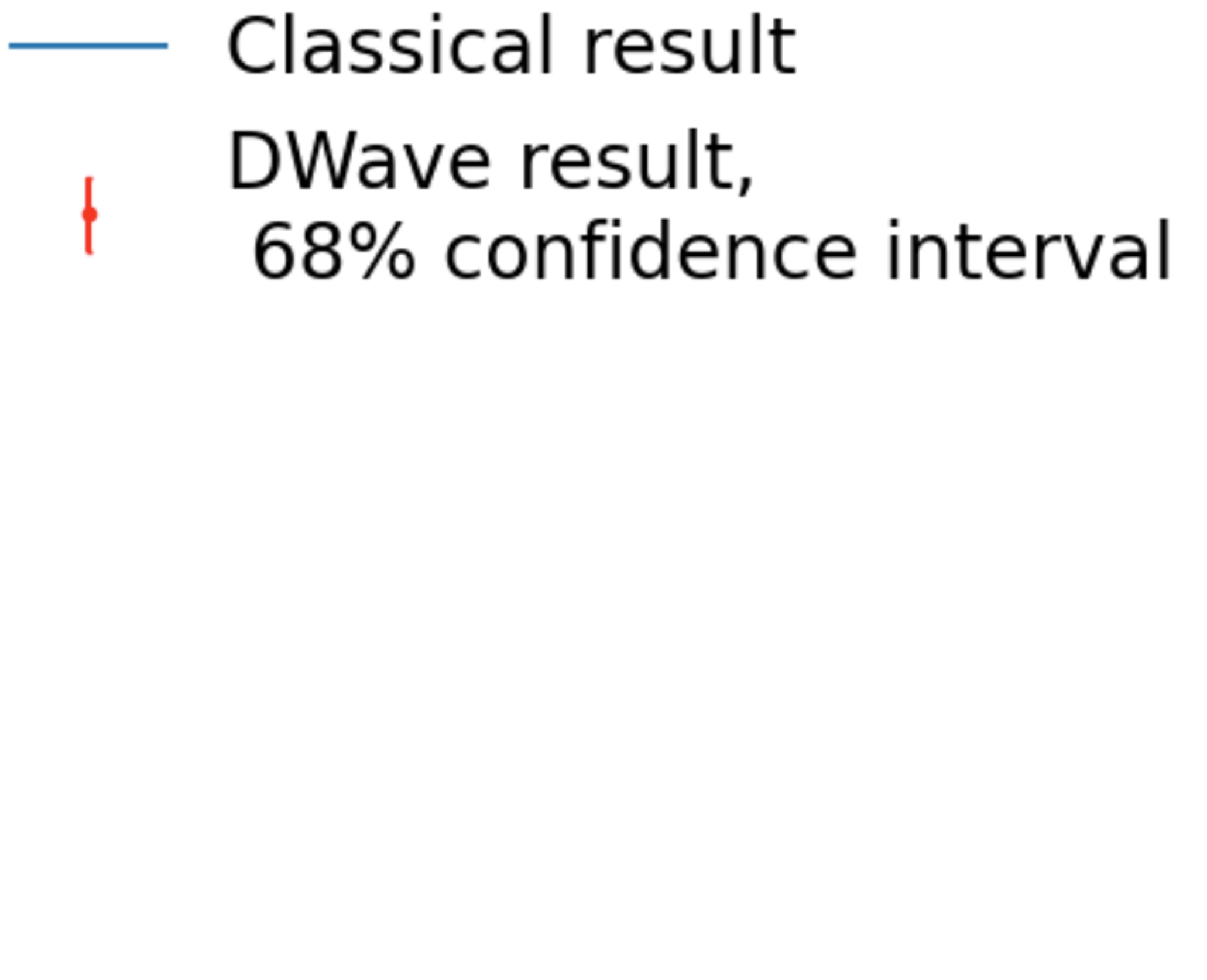}
    \includegraphics[width=0.32\textwidth]{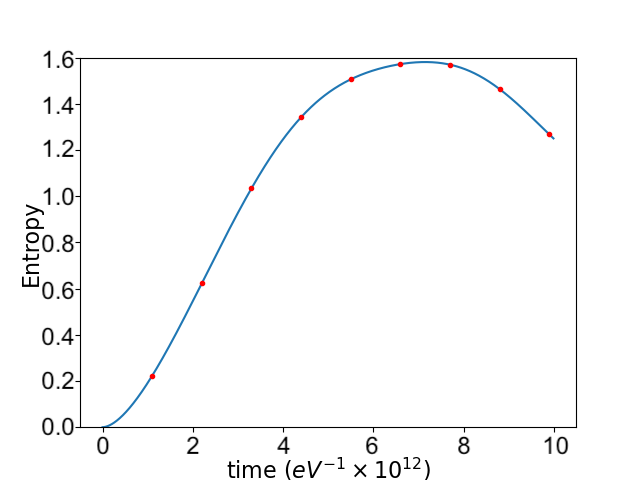} \\

    \includegraphics[width=0.32\textwidth]{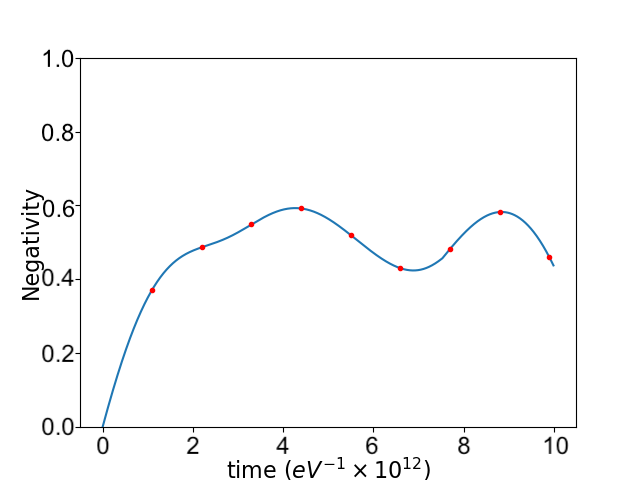}
    \includegraphics[width=0.32\textwidth]{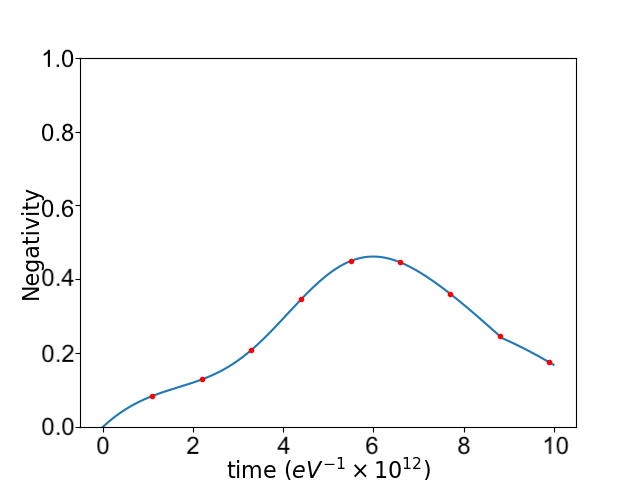}
    \includegraphics[width=0.32\textwidth]{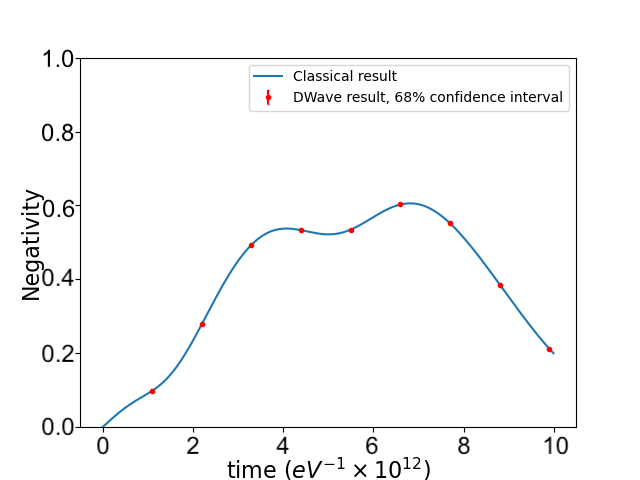}
    
    \caption{A comparison of an N = 4, $n_f$ = 3 collective neutrino oscillation results from {\tt Advantage} and exact classical computation for entanglement entropy of the third (``probe'') neutrino (top), negativity between neutrinos 1 and 3 (bottom left), negativity between neutrinos 2 and 3 (bottom middle), and negativity between neutrinos 3 and 4 (bottom right). The initial state is $\ket{\nu_{e} \nu_{e} \nu_{\tau} \nu_{\mu}}$ and the Hamiltonian parameters are given by Tab. \ref{tab:2flavexactsimulationpars}. The blue icon at the top-right indicates simulation on a quantum device \cite{Klco2020}.} 
    \label{fig:fourneutrinodwavetoclassicalcomparison}
\end{figure}

Next, the results for an initial state of $\ket{\nu_{e} \nu_{e} \nu_{\tau} \nu_{\mu}}$ with Hamiltonian parameters in Tab. \ref{tab:2flavexactsimulationpars} were replicated on {\tt Advantage}. Since only the neutrino-neutrino interaction term of the Hamiltonian in Eq. \ref{eq:H_collectiveneutrinooscilation} affects the time-evolution of the entanglement witnesses, the choice of $\vec{B}(E)$ is arbitrary for this step. Nine different times were sampled, and the following procedure was used for each time: The initial state would be transformed into the mass-basis and split into blocks, each of which is composed of all mass-basis states with a given number of neutrinos in each mass-basis eigenstate. For N = 4 and $n_f$ = 3, there are 3 blocks of size 1, 6 blocks of size 4, 3 blocks of size 6, and 3 blocks of size 12. Time-evolution using the techniques outlined in Sec. \ref{sec:dwaveimplementationdescription} would be done separately for each block. In the event of a nonconvergence, defined as the previous 8 iterations achieving a percentage difference of less than 1\% between the expectation values of the Feynman clock Hamiltonian obtained from Eqs. \ref{eq:QFeynman_AQAEforwardorder} and \ref{eq:QFeynman_AQAEreverseorder} from adjacent iterations, the digitization would be rewound to the last iteration at which the progression of the Feynman clock expectation value was consistent with that of converging samples and would begin again from that point. The results from each block would be put back together to form the result for the state-vector, from which the entanglement entropy and negativity would then be extracted. The results for the ``probe'' third neutrino are seen in Fig. \ref{fig:fourneutrinodwavetoclassicalcomparison}. 
All results are enumerated in Tables \ref{tab:N4nunuexactresults_S3} and \ref{tab:N4nunuexactresults_N3} of App. \ref{app:n4entanglementtabs}.  
Successful convergence to the classical result was achieved on {\tt Advantage} to classical machine precision without Trotter error, similar to what was accomplished in Ref. \cite{illa2022basic}. In addition, the natural block-diagonalization of the neutrino mass basis was demonstrated to be a promising avenue for extending the reach of quantum simulation of $\nu$-$\nu$ interaction.

\section{Other systems}
\label{sec:othersystems}

In the literature discussed in Section \ref{sec:intro} surrounding neutrinos in the early Universe, neutron star mergers, and core-collapse supernovae, antineutrinos play as important of a role as neutrinos.  To address the part that antineutrinos play,  I discuss a method presented in Ref.  \cite{Balantekin2007} for adapting Eq. \ref{eq:H_collectiveneutrinooscilation} to interactions between neutrino and antineutrino modes and produce a few of its notable results for $n_f$ = 3 in an exact classical simulation of a neutrino-antineutrino pair (N = 2).  Although recent literature utilizing mean-field theory methods have found that non-trivial effects in neutrino-antineutrino mixtures require at least N = 3 \cite{Padilla-Gay2022,  Fiorillo2023a,  Fiorillo2023b, Dasgupta2018},  N = 2 is chosen as this sub-project is intended primarily as a trailhead for studies of the full $n_f$ = 3 Hilbert space.  As in Section \ref{sec:collneutoscham}, only coherent elastic forward scattering is taken into account. 

All of the work discussed so far assumes that neutrinos are Dirac fermions. However, there are theories that stipulate that neutrinos are either Majorana fermions or have a Majorana mass \cite{grotz1990weak}. Recent studies have discussed the effects of spin-flip, magnetic moment, and other non-standard neutrino interactions on Majorana neutrinos \cite{Cirigliano2015,  kharlanov2021effects} and have explored the possibility of using astrophysical phenomena such as white dwarfs \cite{adhikary2023neutrino} and neutron stars \cite{alok2022neutron}. To see if Majorana effects are visible for entanglement observables, N = 2 simulations analogous to the ones for neutrino-antineutrino pairs are done for Majorana fermions.

Finally,  since collective neutrino oscillations are time-dependent in many of their physical applications,  the methods and challenges in the way of implementing time-dependent collective neutrino oscillations on {\tt{Advantage}} are also discussed.

\subsection{Neutrino-antineutrino interactions}

\label{sec:nuantinuoscillations}


Eq. \ref{eq:H_collectiveneutrinooscilation} can be re-written in terms of neutrino creation and annihilation operators. The neutrino creation operator can then be set equal to the antineutrino annihilation operator and vice versa, taking advantage of charge-parity-time {\it (CPT)} invariance. These steps produce an s-channel interaction between a neutrino and an antineutrino. Such s-channel interactions only happen for neutrino-antineutrino pairs of the same flavor and each pair is equally likely to produce each other pair, so for $n_f = 3$ the corresponding term in the Hamiltonian is proportional to this matrix:

\begin{equation}
    \bordermatrix{ & \ket{\nu_e \nu_e} & \ket{\nu_e \nu_\mu} & \ket{\nu_e \nu_\tau} & \ket{\nu_\mu \nu_e} & \ket{\nu_\mu \nu_\mu} & \ket{\nu_\mu \nu_\tau} & \ket{\nu_\tau \nu_e} & \ket{\nu_\tau \nu_\mu} & \ket{\nu_\tau \nu_\tau} \cr
      \ket{\nu_e \nu_e} & 1 & 0 & 0 & 0 & 1 & 0 & 0 & 0 & 1 \cr
      \ket{\nu_e \nu_\mu} & 0 & 0 & 0 & 0 & 0 & 0 & 0 & 0 & 0 \cr
      \ket{\nu_e \nu_\tau} & 0 & 0 & 0 & 0 & 0 & 0 & 0 & 0 & 0 \cr
      \ket{\nu_\mu \nu_e} & 0 & 0 & 0 & 0 & 0 & 0 & 0 & 0 & 0 \cr
      \ket{\nu_\mu \nu_\mu} & 1 & 0 & 0 & 0 & 1 & 0 & 0 & 0 & 1 \cr
      \ket{\nu_\mu \nu_\tau} & 0 & 0 & 0 & 0 & 0 & 0 & 0 & 0 & 0 \cr
      \ket{\nu_\tau \nu_e} & 0 & 0 & 0 & 0 & 0 & 0 & 0 & 0 & 0 \cr
      \ket{\nu_\tau \nu_\mu} & 0 & 0 & 0 & 0 & 0 & 0 & 0 & 0 & 0 \cr
      \ket{\nu_\tau \nu_\tau}  & 1 & 0 & 0 & 0 & 1 & 0 & 0 & 0 & 1 }
      \label{eq:schannelhammatrix}
\end{equation}

The matrix in Eq. \ref{eq:schannelhammatrix} is proportional to $(\vec{\lambda}^{*} \otimes I) \cdot (I \otimes \vec{\lambda})$, where $\vec{\lambda}$ is the 8-term vector composed of the Gell-Mann matrices and $I$ is the 3x3 identity matrix. The superscript asterisk denotes complex conjugation. The coefficient multiplied into the matrix in Eq. \ref{eq:schannelhammatrix} is the same as the coefficient of the neutrino-neutrino term in Eq. \ref{eq:H_collectiveneutrinooscilation_nf2}, with the exception of an additional minus sign \cite{notzold_neutrino_1988, sigl_general_1993, Balantekin2007,  Cirigliano2015}. This is because because the Feynman diagrams for neutrino-antineutrino interactions are odd permutations of those for neutrino-neutrino interactions \cite{srednicki2007}. The resulting additional term for neutrino-antineutrino interactions is as follows:

\begin{equation}
    H_{\nu \bar{\nu}} = -\sum_{p,p^{'}} k (1 - \cos{\theta_{p p^{'}}}) \vec{\lambda}^{*}_p \cdot \vec{\lambda}_{p^{'}}
    \label{eq:nuantinuhamterm}
\end{equation}

\noindent where $p^{'}$ must index an antineutrino mode if $p$ indexes a neutrino mode and must index a neutrino mode if $p$ indexes an antineutrino mode. For $n_f = 2$, the s-channel interaction Hamiltonian term is proportional to this matrix:  

\begin{equation}
    \bordermatrix{ & \ket{\nu_e \nu_e} & \ket{\nu_e \nu_\mu} & \ket{\nu_\mu \nu_e} & \ket{\nu_\mu \nu_\mu} \cr
      \ket{\nu_e \nu_e} & 1 & 0 & 0 & 1 \cr
      \ket{\nu_e \nu_\mu} & 0 & 0 & 0 & 0 \cr
      \ket{\nu_\mu \nu_e} & 0 & 0 & 0 & 0 \cr
      \ket{\nu_\mu \nu_\mu} & 1 & 0 & 0 & 1 }
      \label{eq:schannelhammatrixnf2}
\end{equation}

\noindent The matrix in Eq. \ref{eq:schannelhammatrixnf2} is directly proportional to $(\vec{\sigma}^{*} \otimes I) \cdot (I \otimes \vec{\sigma})$, where $\vec{\sigma}$ is the 3-term vector composed of the Pauli matrices and $I$ is the 2x2 identity matrix. Just as for $n_f = 3$, the superscript asterisk denotes complex conjugation. Aside from this change of matrix the process for deriving the $n_f = 2$ neutrino-antineutrino interaction Hamiltonian is the same as for $n_f = 3$ and the term ends up being

\begin{equation}
    H_{\nu \bar{\nu}} (n_f = 2) = -\sum_{p,p^{'}} k (1 - \cos{\theta_{p p^{'}}}) \vec{\sigma}^{*}_p \cdot \vec{\sigma}_{p^{'}}
    \label{eq:nuantinuhamtermnf2}
\end{equation}

In principle, it is also possible for a neutrino and an antineutrino to exchange momentum through the t-channel.  However, a recent study found that such interactions are helicity-suppressed. That is, they are nonexistent in the limit of massless neutrinos and in the physical reality are negligibly small given the small mass of neutrinos \cite{fiorillo2024collective}.  S-channel interactions where the modes change between neutrino and antineutrino are suppressed in the same way.  Hence, Eq. \ref{eq:nuantinuhamterm} fully encapsulates a neutrino-antineutrino system's interactions to a very good approximation.

One marked difference between the neutrino-neutrino and neutrino-antineutrino systems is while the 2-body interaction for the former is independent of the basis that the neutrinos are in for both $n_f$=2 and for $n_f$ = 3,  the 2-body interaction for for the former is only basis-independent for $n_f$=2. Hence, the full Hamiltonian for the neutrino-antineutrino term in the $n_f$=3 case must be written with the single-neutrino terms in the flavor basis, like so:
\begin{equation}
    H^{\nu \bar{\nu}}_{CNO} = \sum_{p = 1}^N {U_{PMNS}}_p \vec{B}(E) \cdot \vec{\lambda}_p  {U_{PMNS}^\dagger}_p -\sum_{p,p^{'}} k (1 - \cos{\theta_{p p^{'}}}) \vec{\lambda}^{*}_p \cdot \vec{\lambda}_{p^{'}}
    \label{eq:nuantinuhamtermfull_nf3}
\end{equation}

\noindent with ${U_{PMNS}}_p$ and ${U_{PMNS}^\dagger}_p$ representing the PMNS matrix applied to the neutrino or antineutrino mode indexed by $p$. As in Sec. \ref{sec:fourneutrinosim}, the energies of the neutrinos are all held to the same value $E$. Additionally, the 2-body interaction commutes with the single-neutrino oscilation term for $n_f$ = 2 but not for $n_f = 3$, so the reasoning behind the treatment of $\vec{B}(E)$ as arbitrary in Sec. \ref{sec:fourneutrinosim} only holds for $n_f$ = 2.  Hence, for $n_f$ = 3, $\vec{B}(E)$ is fixed to $(0, 0, -\frac{\delta m^2}{4 E}, 0, 0, 0, 0, -\frac{\Delta m^2}{2\sqrt{3} E})$, the value obtained in App. \ref{app:singlenuosc}. The $n_f = 2$ counterpart to Eq. \ref{eq:nuantinuhamtermfull_nf3} is

\begin{equation}
    H^{\nu \bar{\nu}}_{CNO} (n_f = 2)= \sum_{p = 1}^N \begin{pmatrix} -\frac{\Delta m^2}{4 E} \cos{2 \theta} & \frac{\Delta m^2}{4 E} \sin{2 \theta} \\ \frac{\Delta m^2}{4 E} \sin{2 \theta} & \frac{\Delta m^2}{4 E} \cos{2 \theta} \end{pmatrix}_p -\sum_{p,p^{'}} k (1 - \cos{\theta_{p p^{'}}}) \vec{\sigma}^{*}_p \cdot \vec{\sigma}_{p^{'}}
    \label{eq:nuantinuhamtermfull_nf2}
\end{equation}

\noindent in the flavor basis, with $\theta$ being the mixing angle that maps between the flavor and mass bases for $n_f = 2$ via an SO(2) transformation on each neutrino.

Because the interaction term represented by Eq. \ref{eq:nuantinuhamterm} is not basis-independent and is specific to the flavor basis, the domain-decomposition technique used in Sec. \ref{sec:fourneutrinosim} to place interacting N = 4 neutrino systems on {\tt Advantage} cannot be used for neutrino-antineutrino systems. This limits the reach of quantum annealers in simulations of mixtures of neutrinos and antineutrinos. 

Fig. \ref{fig:majorana_nuantinu_2flav_vs3flav} shows a comparison of results for the time evolution of entanglement entropy obtained using exact linear algebra on a classical computer for four systems with N = 2 neutrinos, one for each of the following initial states: $\ket{\nu_e \nu_e}$, $\ket{\nu_e \nu_\mu}$, $\ket{\nu_e \bar{\nu}_e}$, $\ket{\nu_e \bar{\nu}_\mu}$. 
Each two-body system is analyzed for both $n_f$ = 2 and $n_f$ = 3. The neutrino-neutrino systems show no difference between the $n_f$ = 2 and $n_f$ = 3 results. However, for the neutrino-antineutrino systems, there is a difference. For $n_f$ = 2,  the same periodic pattern that appears in the initially $\ket{\nu_e \nu_\mu}$ system appears in neutrino-antineutrino system, but with a lower amplitude and half the frequency.  For $n_f$ = 3, the neutrino-antineutrino entanglement entropies are more irregular. The $\ket{\nu_e \bar{\nu}_\mu}$ system in particular shows non-periodic behavior, reminiscent of the Second Law of Thermodynamics, whereby there is a general trend towards the maximum entropy of $log_2(3)$. Changing the physics to set all mixing angles to zero, or setting all mixing angles but $\theta_{v12}$ and one of the mass-difference parameters to 0, are the only ways to make it so that the $n_f$ = 3 neutrino-antineutrino result replicates the $n_f$ = 2 neutrino-antineutrino result.
Physically, this suggests that another neutrino flavor inaccessible through neutrino oscillation but accessible through neutral current would be detectable through neutrino-antineutrino interactions as long as the mass of the new eigenstate does not equal the mean of the previous eigenstates' masses.

A similar analysis, shown in Fig. \ref{fig:majorana_nuantinu_2flav_vs3flav_2bodyonly}, was done for the case where only the neutrino-neutrino and neutrino-antineutrino interaction was kept, which would be a good approximation for the case immediately at the neutrinosphere, the part of the proto-neutron star from where the neutrinos stream into the outer layers from \cite{Bethe1979}. In this scenario, only initial states of $\ket{\nu_e \nu_\mu}$ and $\ket{\nu_e \bar{\nu}_e}$ produce entanglement. For $n_f$ = 2,  the entanglement dynamics are the same as for the neutrino-neutrino case.  $\ket{\nu_e \bar{\nu}_e}$'s entanglement oscillations for $n_f$ = 3 have $\frac{3}{4}$ of the frequency of, experience a bimodal character not present in, and have a higher maximum ($log_2 (3)$ vs 1) than corresponding oscillations for $n_f$ = 2. This $n_f$ = 3 bimodal behavior may be what manifests in more complex behavior when the single-neutrino oscillation is present.  However, the nonperiodic character of the $\ket{\nu_e \bar{\nu}_\mu}$ seems to be dependent on the existence of vacuum neutrino oscillation effects.

\subsection{Majorana Neutrinos}
\label{sec:majorananu}

As formulated by Steven Weinberg in the 1970s, the five-dimensional Standard Model can include the following term:
\begin{equation}
    \mathcal{L}_{Weinberg} = f_{abmn} \bar{l}^{C}_{i a L} l_{j b L} \phi^{(m)}_{k} \phi^{(n)}_{l} \epsilon_{ik} \epsilon_{jl} + f_{abmn}^{'} \bar{l}^{C}_{i a L} l_{j b L} \phi^{(m)}_{k} \phi^{(n)}_{l} \epsilon_{ij} \epsilon_{kl}
    \label{eq:weinberglagrangian}
\end{equation}

\noindent where $f_{abmn}$ and $f_{abmn}^{'}$ are coefficients with a magnitude on the order of $\frac{1}{M}$ with M being a characteristic mass above approximately $10^{14}$ GeV of hypothetical 'superheavy' particles, $l_{j b L}$ is the $j^{th}$ component of the vector in the fundamental representation of SU(2) representing a non-conserved lepton (in this case, a neutrino) with flavor $b$ with left-handed chirality, the superscript ${C}$ denotes charge-conjugation, and $\phi^{(m)}_{k}$ is a scalar field that the Higgs field can be substituted in for \cite{weinberg1979baryon}. On a staggered lattice, where each spatial lattice site $i$ is split into two sites, $l$ for neutrinos and $l^{'}$ for antineutrinos, $\mathcal{L}_{Weinberg}$ can be converted into a Hermitian Hamiltonian form, like so:

\begin{equation}
    H^{Majorana}_{m} = \frac{1}{2} m_M \sum_{i} (a_{i l} a_{i l^{'}} + a^\dagger_{i l} a^\dagger_{i l^{'}})
    \label{eq:staggeredlatticce}
\end{equation}

\noindent with $a$ and $a^{\dagger}$ being creation and annihilation operators for the neutrino or antineutrino that exists on the staggered lattice site denoted by its subscript, with the creation operator for neutrinos being the annihilation operator for antineutrinos and vice versa. This term produces a ``Majorana mass'' $m_M$ for the neutrinos \cite{Farrell2023b}.

For a single-spatial-site lattice localized to spatial site $i$, the eigenstate of $H^{Majorana}_{m}$ which has a ``mass'' of $m_M$ is $\frac{\ket{\nu} + \ket{\bar{\nu}}}{\sqrt{2}}$, that is, an equal superposition of there being one neutrino on the spatial lattice site and there being one antineutrino on the spatial lattice site. This state can be interpreted as the projection of a Majorana neutrino onto the Dirac neutrino state-space used so far in this work. From this fact, one can obtain the interaction of any pair of Majorana neutrinos through direct addition of the two-neutrino term in Eq. \ref{eq:H_collectiveneutrinooscilation} and its counterpart for antineutrino-antineutrino interactions (which is the same \cite{Balantekin2007}) to the neutrino-antineutrino interaction term in Eq. \ref{eq:nuantinuhamterm}. In other words, $H^{majorana}_{\nu \nu} = H^{Dirac}_{\nu \nu} + 2 H^{Dirac}_{\nu \bar{\nu}} + H^{Dirac}_{\bar{\nu} \bar{\nu}}$. The single-neutrino oscillation term is the same as in the Dirac case. The resulting collective neutrino oscillation Hamiltonian for Majorana neutrinos is as follows:

\begin{equation}
    H^{majorana}_{\nu \nu} = \sum_p {U_{PMNS}}_p \vec{B}(E) \cdot \vec{\lambda}_p  {U_{PMNS}^\dagger}_p + 2 \times \sum_{p,p^{'}} k (1 - \cos{\theta_{p p^{'}}}) Im(\vec{\lambda}_p) \cdot Im(\vec{\lambda}_{p^{'}})
    \label{eq:H_collectiveneutrinooscilationmajorana}
\end{equation}

for $n_f = 3$ and

\begin{equation}
    H^{majorana}_{\nu \nu} (n_f = 2) = \sum_p \begin{pmatrix} -\frac{\Delta m^2}{4 E} \cos{2 \theta} & \frac{\Delta m^2}{4 E} \sin{2 \theta} \\ \frac{\Delta m^2}{4 E} \sin{2 \theta} & \frac{\Delta m^2}{4 E} \cos{2 \theta} \end{pmatrix}_p + 2 \times \sum_{p,p^{'}} k (1 - \cos{\theta_{p p^{'}}}) Im(\vec{\sigma}_p) \cdot Im(\vec{\sigma}_{p^{'}})
    \label{eq:H_collectiveneutrinooscilationmajorana_nf2}
\end{equation}

\noindent for $n_f = 2$. Both Eq. \ref{eq:H_collectiveneutrinooscilationmajorana} and \ref{eq:H_collectiveneutrinooscilationmajorana_nf2} are in the flavor basis and Im() denotes the imaginary component. As in Sec. \ref{sec:fourneutrinosim} and \ref{sec:nuantinuoscillations}, the energies of the neutrinos are all held to the same value $E$. In the same manner as with the neutrino-antineutrino oscillation, the two-body interaction term is not invariant with a change of basis for $n_f = 3$ (though it is for $n_f = 2$), so $\vec{B}(E)$ is fixed to the same value as in Sec. \ref{sec:nuantinuoscillations}. 

\begin{figure}
    \raggedleft
    \includegraphics[width=0.05\textwidth]{ideal-quantum-simulation-classical-hardware.png}

    \centering
    \includegraphics[width=0.4\textwidth]{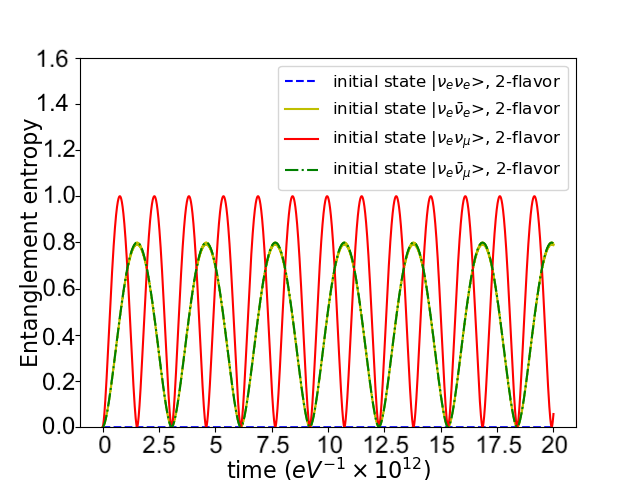}
    \includegraphics[width=0.4\textwidth]{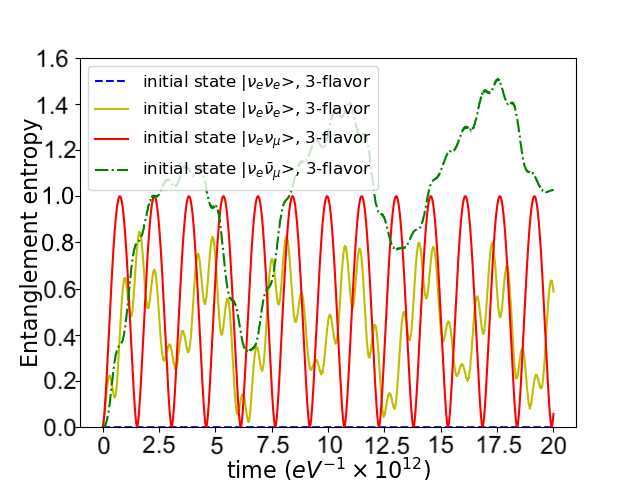} \\
    \includegraphics[width=0.4\textwidth]{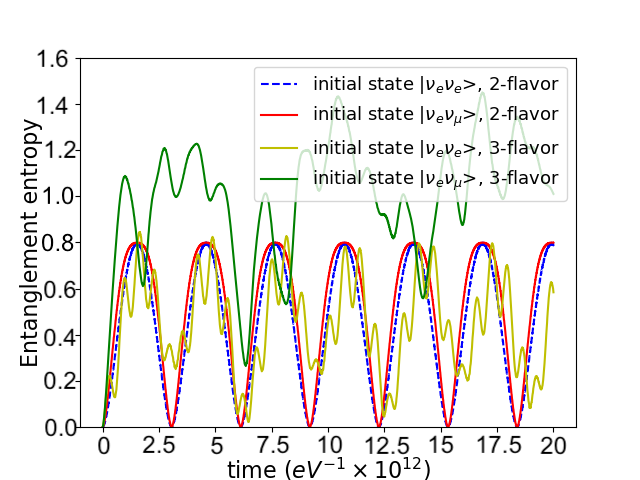}
    \caption{Top: A comparison, which assumes the neutrinos are Dirac particles, of 2-neutrino and 1 neutrino-1 antineutrino (both N = 2) systems' entanglement entropy evolution. $n_f$ = 2 is in the left, $n_f$ = 3 is on the right. Bottom: an analogous comparison for Majorana neutrinos. Parameters are shown in Tab. \ref{tab:2flavexactsimulationpars}. The purple icon at the top-right indicates exact simulation on a classical device \cite{Klco2020}.}
    \label{fig:majorana_nuantinu_2flav_vs3flav}
\end{figure}

Fig. \ref{fig:majorana_nuantinu_2flav_vs3flav} shows results for Majorana neutrinos.  {\it A priori}, the Majorana $\ket{\nu_e \nu_e}$ state should behave similarly to the Dirac $\ket{\nu_e \bar{\nu}_e}$ and the Majorana $\ket{\nu_e \nu_\mu}$ should behave similarly to the Dirac $\ket{\nu_e \nu_\mu}$, and several results agree with this.  First, the $n_f$ = 3 $\ket{\nu_e \nu_\mu}$ initial state case shows a similar pattern of irregularity to its Dirac counterpart.  Second, the Majorana $\ket{\nu_e \nu_e}$ exhibits the same behavior as does the Dirac $\ket{\nu_e \bar{\nu}_e}$  Third,  changing the physics to set all mixing angles to zero, or to set all mixing angles but $\theta_{v12}$ and one of the mass-difference parameters to 0, are the only ways to make it so that the $n_f$ = 2 results replicate the $n_f$ = 3 results.  One difference is that lower-amplitude, higher-frequency modes in the $n_f$ = 3 $\ket{\nu_e \nu_\mu}$ initial state case have a higher prominence for Majorana than for the Dirac neutrinos.
Additionally, the frequency of the oscillation for periodic patterns in the Majorana is half that of counterparts in the Dirac neutrino-neutrino case.  If these differences persist at asymptotically large N,  neutrinos from the initial burst at the collapse of an iron core, which are probably composed primarily of electron neutrinos, could potentially be used to probe whether neutrinos are Majorana or Dirac fermions.

\begin{figure}
    \raggedleft
    \includegraphics[width=0.05\textwidth]{ideal-quantum-simulation-classical-hardware.png}
    
    \centering
    \includegraphics[width=0.4\textwidth]{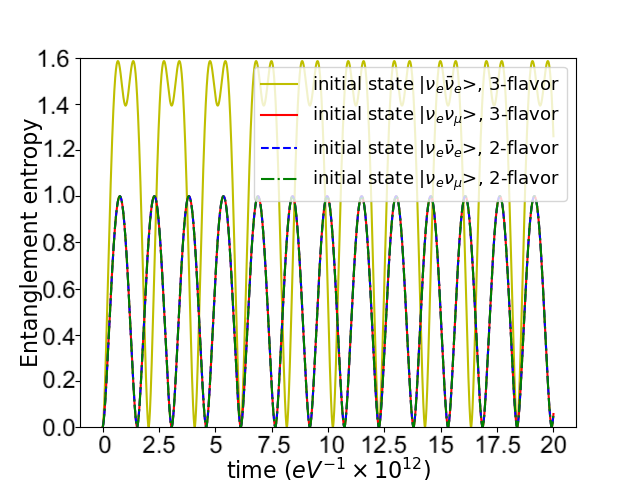}
    \includegraphics[width=0.4\textwidth]{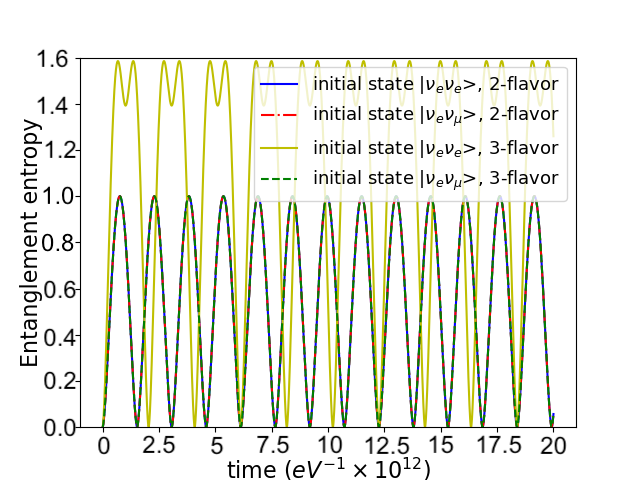}
    \caption{A comparison of Dirac and Majorana N= 2 neutrino systems' entanglement entropy evolution for $n_f$ = 2 and $n_f$ = 3, representative of all scenarios that produce nonzero results. Parameters are the same as those in Tab. \ref{tab:2flavexactsimulationpars}. The purple icon at the top-right indicates exact simulation on a classical device \cite{Klco2020}.}
    \label{fig:majorana_nuantinu_2flav_vs3flav_2bodyonly}
\end{figure}

Similarly to Sec.  \ref{sec:nuantinuoscillations},  the dynamics were also sampled for the case where only the collective neutrino oscillation term is included in the Hamiltonian, and the results, presented in Fig. \ref{fig:majorana_nuantinu_2flav_vs3flav_2bodyonly}, are the same as those of their Dirac counterparts. Thus,  vacuum neutrino oscillation effects in dense neutrino environments are likely an explanation for the differences in the entanglement dynamics of Dirac and Majorana collective neutrino oscillations.


\subsection{Time-dependent Hamiltonian}
\label{sec:timedependentHamsimulation}

As mentioned at the beginning of Section \ref{sec:collneutoscham}, time-dependence of neutrino-neutrino interaction, despite being physical, is excluded from this study in order to facilitate algorithm-development on the quantum annealer. With a time-independent Hamiltonian, expession of the time-evolution in the form of a QUBO matrix is a straightforward use of the time-evolution operator, $e^{-i H t}$. However, for a time-dependent Hamiltonian, the time-evolution operator is the time-ordered integral \cite{srednicki2007},
\begin{equation}
    T(e^{-i \int_{t_i}^{t_f} H(t) dt})
    \label{eq:timedeptimeev}
\end{equation}

\noindent Thus, the use of the same techniques with a time-dependent Hamiltonian would result in an error in the calculation determined by a continuum limit of the Zassenhaus formula \cite{Zassenhaus_Unpublished, Magnus_1954},

\begin{equation}
    e^{t (X+Y)} = e^{t X} e^{t Y} e^{-\frac{t^2}{2} [X,Y]} e^{-\frac{t^3}{6} (2[Y,[X, Y]] + [X,[X, Y]])} e^{-\frac{t^4}{24} ([[[X,Y],X],X] + 3[[[X,Y],X],Y] + 3[[[X,Y],Y],Y])} \cdots
    \label{eq:zassenhausfla}
\end{equation}

\noindent which removes one of the main advantages of using a quantum annealer, which is the avoidance of Trotter errors. It is possible to avoid these Trotter errors, but that would require the Schr\"{o}dinger equation to be fully solved in order to represent the time evolution in QUBO form, which is sufficient to find the time evolution by itself. 


\section{Conclusion}
\label{sec:conclusion}

This work has presented the time-evolution of the entanglement structure of multi-angle, three-flavor coherent collective neutrino oscillations on a quantum annealer. Successful verification of annealer results and extension to neutrino-antineutrino and Majorana neutrino pair dynamics using exact numerical calculations have also been discussed.  Following the direction of Ref. \cite{illa2022basic}, the Feynman Clock method was used to encode the coherent collective neutrino oscillation dynamics onto a Hamiltonian whose ground state emulates the desired time-evolution. The Adaptive Quantum Annealing Eigensolver (AQAE) method was then used to obtain said ground state. Fitting N = 4 neutrinos that each have $n_f$ = 3 possible flavors onto the limited number of logical qubits available required the use of domain-decomposition.  This was accomplished by taking advantage of the fact that in the mass-basis, the collective neutrino oscillation Hamiltonian is naturally block-diagonalized. This technique is only possible with a neutrino-neutrino interaction that is invariant with respect to neutrino-basis transformation under the Pontecorvo–Maki–Nakagawa–Sakata (PMNS) matrix. Therefore, while it works for systems of Dirac neutrinos, it does not work for neutrino-antineutrino mixtures or for Majorana neutrinos. 


AQAE can reproduce collective neutrino oscillation dynamics without Trotter errors for $n_f$ = 3,  just like how it was able to for $n_f$ = 2 in Ref. \cite{illa2022basic},  and do so with a precision equal to the maximum attainable with the double-precision floating point format on classical computers.  However, its qubit-requirements do not scale any better than do those of classical devices. 
This obstacle most likely cannot be overcome in the general case. This is because $H_f$, as defined in Sec. \ref{sec:dwaveimplementationdescription}, is a stoquastic Hamiltonian (that is, all of its off-diagonal elements are real and non-positive). The problem of solving the ground state of a stoquastic Hamiltonian can be mapped onto classical statistical problems \cite{bravyi2006complexity, bravyi2017complexity}. Replicating this mapping for a general-case quantum simulator would be equivalent to the dequantization of the complexity class QMA \cite{bravyi2006complexity}. Thus, such an achievement, and by extension the mapping of a general-case quantum simulator to a stoquastic Hamiltonian ground state problem, is highly unlikely. Thus in the general case it will likely take annealers with alternative choices of $H_\alpha$ (in Sec.  \ref{sec:dwaveimplementationdescription}'s parlance), such as the ZZXX and the ZX Hamiltonian \cite{aharonov2008adiabatic, biamonte2008realizable} to efficiently conduct quantum computation in the general case. In principle, it is possible to make {\tt Advantage} into a universal quantum computer through measurement-based quantum computing (MBQC). This is because $H_f$, as defined in Sec. \ref{sec:dwaveimplementationdescription}, can be used to obtain a cluster state for MBQC from $H_i$'s (again, as defined in Sec. \ref{sec:dwaveimplementationdescription}) ground state \cite{raussendorf2003measurement}, so applying MBQC on {\tt Advantage} can theoretically be done via quenches from $H_i$ to $H_f$. One challenge that this will entail is that many implementations of such a transformation of {\tt Advantage}'s function will involve the sacrifice of the noise-resilience that quantum annealing has in the form of a straightforward measure of the quality of an anneal's result in the form of $H_f$'s expectation value. This brings us to another difficulty that {\tt Advantage} faces, which is its high level of noise. This is connected to {\tt Advantage}'s low decoherence time of $\approx$30 ns \cite{king2023quantum}, which is shorter than even the minimum anneal-time, 100 ns, available on the device  \cite{dwavesysdocs}. The result of such extensive noise is that the procedure outlined in Sec. \ref{sec:fourneutrinosim} and App. \ref{app:nealbenchmarking} requires hundreds of anneals and maximization of the magnitude of the penalty term in order to produce one anneal whose result is sufficiently accurate to enable AQAE to converge. Furthermore, even with these improvements convergence was not achieved for Hilbert spaces composed of 30 or more states.

D-Wave Inc.'s new {\tt Advantage2} annealer has more qubits, longer coherence time ($\approx$300 ns compared to {\tt Advantage's} $\approx$30 ns), and less noise than {\tt Advantage }\cite{amin2023quantum}. Thus, it is possible that paths toward quantum advantage on annealers could be explored on {\tt Advantage2} and its successors.  However, given the fact that universal gate-based devices with thousands of qubits, such as such as IBM's Condor processor, exist and are expected to grow in capability \cite{Gambetta_2023}, gate-based devices look to be more promising than annealers for realizing quantum advantage for quantum simulations in general.  Quantum annealers will most likely be more suited towards specialization in problems that can be mapped to a search for the ground state of stoquastic Hamiltonians, such as optimization problems.  Recent studies have found evidence that quantum annealing and classical simulation of quantum annealing achieve equal to superior performance over conventional classical algorithms \cite{albash2018demonstration,  bauza2024scaling, ismail2024quantum,fornetti2024solving} and gate-based quantum approaches \cite{mcgeoch2024comment} for several optimization problems.  


For the case of a pure Dirac neutrino sample without antineutrinos, I have found that the entanglement dynamics of neutrinos with the physical $n_f$ = 3 are different from what can be produced by an $n_f$ = 2 system if and only if all three neutrino flavors are present in the initial state. This suggests that for a near purely $\nu_e$ sample,  which is likely a close approximation of the composition of the neutrinos first emitted as the iron core of a massive star collapses \cite{fischer_neutrino_2012, fischer_protoneutron_2010, huedepohl_neutrino_2010}, relevant three-flavor dynamics can be simulated efficiently on classical computers with methods such as mean field theory, while any entangling phenomena can be obtained using a two-flavor simulation.  This is in contrast to results from Ref. \cite{siwach2023entanglement}, which showed a difference in results between two-flavor and three-flavor models for a system whose initial state is entirely composed of electron neutrinos.  One possible explanation for this apparent discrepancy is that a difference between $n_f$ = 2 and $n_f$ = 3 manifests itself when either the neutrino-neutrino interaction is time-dependent or the neutrino modes have different energies, and investigation of this is a potential topic for future work.


However, if physical neutrinos are Majorana fermions,  the findings of this work suggest a far different outcome. Even for the simple case where the number of neutrinos, N, is 2, Majorana neutrino entanglement shows a marked difference in behavior between $n_f$ = 2 and $n_f$ = 3 for all initial states. This result indicates that the entanglement structure of the initial burst of electron neutrinos from the core-collapse of a massive star can potentially indicate if neutrinos exhibit a Majorana mass or not. Thus, the behavior of collective Majorana neutrino oscillations in larger systems and their effects on supernova observables are a potential next step.  It is important to note that the divergent behavior of Majorana fermions observed in this study is dependent on both vacuum neutrino oscillation and neutrino-neutrino interactions having a non-negligible effect, so the best place to look for this behavior is likely the transition region between the interaction-dominated regime and the vacuum oscillation-dominated regime.

O(1 second) after core collapse, the neutrinos from the core are expected to be a roughly even mixture of all three flavors and of neutrinos and antineutrinos \cite{fischer_neutrino_2012, fischer_protoneutron_2010, huedepohl_neutrino_2010}. For such a system, I have found that $n_f$ = 3 entanglement dynamics are distinct from those for $n_f$ = 2. This is due to two reasons. The first is the difference with $n_f$ = 2 behavior exhibited by a state that starts as a mixture of all three flavors. The second is because, similar to Majorana neutrinos, neutrino-antineutrino mixtures show substantial differences between $n_f$ = 2 and $n_f$ = 3 even for the simple N = 2 case when both neutrino-neutrino interactions and vacuum oscillations have non-negligible effects on flavor dynamics. In conclusion, a simulation of larger-N neutrino-antineutrino mixtures with all three flavors equally represented could be an ideal substrate for realizing quantum advantage.

This study is intended primarily as a demonstration, so there are several directions in which future work could proceed, besides the straightforward expansion of $n_f$ = 3 dynamics to larger-capacity quantum devices. The first is to evaluate the full neutral current neutrino-neutrino interaction and compare its entanglement dynamics to that found in the coherent forward scattering approximation.  The second is to find a method of evaluating neutrino-antineutrino interactions either on quantum annealers or on gate-based devices and to do so for at least three modes (which Refs.  \cite{PhysRevD.97.023017, PhysRevD.107.123024, PhysRevD.107.043024, PhysRevLett.128.121102} found to be the minimum $N$-value for which mean-field instabilities occur).  This could be a first step towards realizing quantum advantage in $n_f$ = 3 environments with an even mixture of neutrinos and antineutrinos of all 3 flavors.
Third, until quantum computers sufficiently large-scale and high-fidelity to simulate collective neutrino dynamics at a macroscopic scale become available,  methods of modifying mean-field theory to successfully model $n_f$ = 3 quantum entanglement effects on astrophysical observables, along with other macroscopic heuristics, will be important to develop.  A fourth direction is repeating for $n_f = 3$ searches for conserved quantities, Loschmidt echoes, thermalization,  dynamical phase transitions, and other phenomena that have been conducted for $n_f = 2$ \cite{pehlivan2011invariants, Espinoza2013, Pehlivan2014, birol2018neutrino, Cervia2019, Cervia2019Entanglement, patwardhan2019eigenvalues, patwardhan2021spectral, roggero2021dynamical, roggero2021entanglement, Lacroix2022, Martin2023ManyBody, cirigliano2024neutrino, Bhaskar2024TimeScales}.

\section*{Acknowledgements}

This work was supported in part by U.S. Department of Energy, Office of Science, Office of Nuclear Physics, InQubator for Quantum Simulation (IQuS) \cite{IQUSwebsite} under Award Number DOE (NP) Award DE-SC0020970 via the program on Quantum Horizons: QIS Research and Innovation for Nuclear Science and by the Quantum Computing Summer School 2023 at Los Alamos National Laboratory (LANL) \cite{LANLQCSSwebsite} which is sponsored by the LANL Information
Science and Technology Institute and by extension
supported by the U.S. Department of Energy.

I acknowledge the use of DWave Systems Inc.'s services for this work \cite{DWavewebsite}. The views expressed are those of the author and do not reflect the official policy or position of DWave Systems. In this paper, I used {\tt Advantage}, DWave Systems's latest device as of the beginning of this project. This project also made extensive use of Python, Wolfram Mathematica, and Julia. Two of the most important libraries used were DWave Systems's Ocean environment and the QuantumAnnealing Julia library, developed by Carleton Coffrin and Zachary Morrell \cite{QuantumAnnealingwebsite}.

I would like to thank Carleton Coffrin and Zachary Morrell from LANL for guiding me through the use of DWave's systems and assisting with access to DWave's machines. I would also like to thank Professor Martin Savage at IQuS and Joseph Carlson at LANL's T-2 group for their mentorship throughout this project,  for creating the idea of a neutrino project on D-Wave Inc.'s devices, and for helping secure access to the {\tt Advantage} annealer.  Additionally, I would like to thank the staff involved in the organization of the Quantum Computing Summer School (QCSS) at LANL, particularly Lukasz Cincio and Marco Cerezo.
I would also like to thank my colleagues from iQuS, Stephan Caspar, Francesco Turro, and Marc Illa,for valuable discussions.  Finally,  I would like to thank my colleagues from the T-2 group,  Joshua Martin, Ionel Stetcu, and Ronen Weiss, both for valuable discussions and for assistance with access to DWave Systems Inc.'s software and devices.

\appendix

\section{Derivation of the $n_f$ = 3 single-neutrino oscillation Hamiltonian term}
\label{app:singlenuosc}

First, assuming that the velocity, and hence the momentum of all 3 mass-eigenstates of the neutrino are the same, and expressing each eigenstate's energy using the relativistic energy formula, the Hamiltonian is as follows:

\begin{equation}
    H_\nu = \sqrt{m_1^2 + p^2} \ket{1} \bra{1} + \sqrt{m_2^2 + p^2} \ket{2} \bra{2} + \sqrt{m_3^2 + p^2} \ket{3} \bra{3}
    \label{eq:firststepH_nu}
\end{equation}

In the ultrarelativistic limit, which neutrinos with energies of 10 MeV can be approximated to be in, one can take the Taylor expansion of the square root function around the zero-mass limit:

\begin{equation}
    H_\nu = p(1 + \frac{m_1^2}{2 p^2}) \ket{1} \bra{1} + p(1 + \frac{m_2^2}{2 p^2}) \ket{2} \bra{2} + p(1 + \frac{m_3^2}{2 p^2}) \ket{3} \bra{3}
    \label{eq:secondstepH_nu}
\end{equation}

Because the zero-point of energy is arbitrary, one can simply subtract the identity matrix times $(p + \frac{m_1^2 + m_2^2 + m_3^2}{6 p^2})$ to make the Hamiltonian traceless:

\begin{equation}
    H_\nu = \frac{2 m_1^2 - m_2^2 - m_3^2}{6 p} \ket{1} \bra{1} +\frac{2 m_2^2 - m_1^2 - m_3^2}{6 p} \ket{2} \bra{2} + \frac{2 m_3^2 - m_1^2 - m_2^2}{6 p} \ket{3} \bra{3}
    \label{eq:thirdstepH_nu}
\end{equation}

Defining $\delta m^2 = m_2^2 - m_1^2$ and $\Delta m^2 = m_3^2 - \frac{m_2^2 + m_1^2}{2}$, one can get

\begin{equation}
    H_\nu = \frac{ - \frac{3}{2} \delta m^2 - \Delta m^2}{6 p} \ket{1} \bra{1} +\frac{\frac{3}{2} \delta m^2 - \Delta m^2}{6 p} \ket{2} \bra{2} + \frac{\Delta m^2}{3 p} \ket{3} \bra{3}
    \label{eq:fourthstepH_nu}
\end{equation}

And given that in the ultrarelativistic limit, $p \simeq E$, 

\begin{equation}
    H_\nu =  (-\frac{\delta m^2}{4 E} - \frac{\Delta m^2}{6 E})\ket{1} \bra{1} + (\frac{\delta m^2}{4 E}- \frac{\Delta m^2}{6 E})\ket{2} \bra{2} + (\frac{\Delta m^2}{3 E})\ket{3} \bra{3}
    \label{eq:fifthstepH_nu}
\end{equation}

Or: 

\begin{equation}
    H_\nu = - \frac{\delta m^2}{4 E} \lambda_3 - \frac{\Delta m^2}{2 \sqrt{3} E} \lambda_8
    \label{eq:sixthstepH_nu}
\end{equation}

\section{AQAE derivation and demonstration}
\label{app:AQAEderivdemonstration}

AQAE and the methods used in its derivation are demonstrated in this appendix using a single-neutrino system evolving under the Hamiltonian in Eq. \ref{eq:fifthstepH_nu} mapped to the flavor basis using the PMNS matrix from Eq. \ref{eq:PMNS3flav_definition}. The parameters in the Hamiltonian and the PMNS matrix are set to those in the second column of Tab. \ref{tab:2flavexactsimulationpars}. In the flavor basis, this Hamiltonian (expressed in units of $10^{-13} eV$) becomes 

\begin{equation}
    H^{flav}_{\nu} = \begin{pmatrix}
        -386.743 & -34.0125 + 125.961 i & -53.2951 + 112.899 i\\
        -34.0125 - 125.961 i & 258.892 & 589.399 -2.39410 i \\
        -53.2951 - 112.899 i & 589.399 + 2.39411 i & 127.851 \\
    \end{pmatrix}
      \label{eq:1nuflavham}
\end{equation}

\noindent Using Eq. \ref{eq:feynmanclockham}, the matrix-form of the two-time-step Feynman clock Hamiltonian for the time-evolution of a system starting as an electron neutrino and evolving by a time of $10^{13} eV^{-1}$ under $H^{flav}_{\nu}$ is 

\begin{equation}
\begin{split}
H^{\text{Feynman}}_{\nu} = \hspace{90ex} \\
\begin{bmatrix}
0.5 & 0 & 0 & 0.4769 e^{-1.2664 i} & 0.1319 e^{3.0781 i} & 0.0718 e^{0.5594 i} \\
0 & 1.5 & 0 & 0.0934 e^{-2.4350 i} & 0.4067 e^{-0.8704 i} & 0.2755 e^{0.6729 i} \\
0 & 0 & 1.5 & 0.1176 e^{-0.2030 i} & 0.2593 e^{0.5439 i} & 0.4110 e^{-0.9836 i} \\
0.4769 e^{1.2664 i} & 0.0934 e^{2.4350 i} & 0.1176 e^{0.2030 i} & 0.5 & 0 & 0 \\
0.1319 e^{-3.0781 i} & 0.4067 e^{0.8704 i} & 0.2593 e^{-0.5439 i} & 0 & 0.5 & 0 \\
0.0718 e^{-0.5594 i} & 0.2755 e^{-0.6729 i} & 0.4110 e^{0.9836 i} & 0 & 0 & 0.5 \\
\end{bmatrix}
\end{split}
\label{eq:1nnu_sampleFCH}
\end{equation}

\noindent The ground state of this Feynman clock Hamiltonian, normalized to a magnitude of $\sqrt{2}$, is 

\begin{equation}
\ket{FCH g.s.} = (1, 0, 0, -0.285902 - 0.909988i, 0.263185 + 0.0167209i, -0.121759 + 0.0762265i)
\label{eq:FCH_groundstate}
\end{equation}

\noindent which, given that 

\begin{equation}
    e^{i H^{flav}_\nu (10^{13} eV^{-1})}\begin{pmatrix}
    1 \\
    0 \\
    0
    \end{pmatrix} = \begin{pmatrix}
    -0.285902 - 0.909988i \\
    0.263185 + 0.0167209i \\
    -0.121759 + 0.0762265i
    \end{pmatrix}
    \label{eq:fchgs_timeevcheck}
\end{equation}

\noindent matches the expectation that the ground state of $H^{\text{Feynman}}_{\nu}$ is a concatenation of the initial state and the state after a time-evolution with $H^{flav}_{\nu}$. 

Both the QUBO model Hamiltonian and the Ising model Hamiltonian, the two native methods of encoding a problem for {\tt Advantage} to solve \cite{dwavesysdocs}, are both real matrices. Thus, in order to be solved on {\tt Advantage}, the Feynman clock Hamiltonian must be mapped to a real matrix. Absent a method of transforming the Feynman clock Hamiltonian into a basis where it is real, the method used in Refs. \cite{Teplukhin2020, illa2022basic} is used. This method works by doubling the size of the Hilbert space and assigning half of the new, oversized Hilbert space to the real parts of a state's state-vector and the other half to the imaginary part of a state's state-vector, like so:

\begin{equation}
    \begin{pmatrix}
    a_1 + i b_1 \\
    a_2 + i b_2 \\
    \cdots \\
    a_n + i b_n
    \end{pmatrix} \rightarrow
    \begin{pmatrix}
    a_1 \\
    \cdots \\
    a_n \\
    \hline
    b_1 \\
    \cdots \\
    b_n
    \end{pmatrix}
    \label{eq:realtoimagmapping_sv}
\end{equation}

\noindent where ${a_1, \ldots, a_n}$ and ${b_1, \ldots, b_n}$ are all real numbers. The corresponding mapping for a Hamiltonian $H$ is as follows:

\begin{equation}
    \begin{pmatrix}
    H_{11} & \ldots & H_{1n} \\
    \vdots & \ddots & \vdots \\
    H_{n1} & \ldots & H_{nn}
    \end{pmatrix} \rightarrow \left( \begin{array}{ccc|ccc}
    Re(H_{11}) & \ldots & Re(H_{1n}) & -Im(H_{11}) & \ldots & -Im(H_{1n}) \\
    \vdots & \ddots & \vdots & \vdots & \ddots & \vdots \\
    Re(H_{n1}) & \ldots  & Re(H_{nn}) & -Im(H_{n1}) & \ldots  & -Im(H_{nn}) \\
    \hline 
    Im(H_{11}) & \ldots & Im(H_{1n}) & Re(H_{11}) & \ldots & Re(H_{1n}) \\
    \vdots & \ddots & \vdots & \vdots & \ddots & \vdots \\
    Im(H_{n1}) & \ldots  & Im(H_{nn}) & Re(H_{n1}) & \ldots  & Re(H_{nn}) \\
\end{array} \right)
    \label{eq:realtoimagmapping_ham}
\end{equation}

\noindent where $Re()$ and $Im()$ denote real and imaginary components, respectively. In this appendix, $H^{Feynman}_{Re}$ refers to the real-valued matrix obtained from applying the transformation in Eq. \ref{eq:realtoimagmapping_ham} to the Feynman clock Hamiltonian as defined by Eq. \ref{eq:feynmanclockham}. For $H^{\text{Feynman}}_{\nu}$ from  Eq. \ref{eq:1nnu_sampleFCH}, the mapping from Eq. \ref{eq:realtoimagmapping_ham} produces

\begin{equation}
\begin{split}
H^{\text{Feynman}}_{Re(\nu)} = \hspace{90ex} \\
\hspace{-2ex} \begin{bmatrix}
0.5 & 0 & 0 & 0.143 & -0.132 & 0.061 & 0 & 0 & 0 & 0.455 & -0.008 & -0.038 \\
0 & 1.5 & 0 & -0.071 & 0.262 & 0.215 & 0 & 0 & 0 & 0.061 & 0.311 & -0.172 \\
0 & 0 & 1.5 & 0.115 & 0.222 & 0.228 & 0 & 0 & 0 & 0.024 & -0.134 & 0.342 \\
0.143 & -0.071 & 0.115 & 0.5 & 0 & 0 & -0.455 & -0.061 & -0.024 & 0 & 0 & 0 \\
-0.132 & 0.262 & 0.222 & 0 & 0.5 & 0 & 0.008 & -0.311 & 0.134 & 0 & 0 & 0 \\
0.061 & 0.215 & 0.228 & 0 & 0 & 0.5 & 0.038 & 0.172 & -0.342 & 0 & 0 & 0 \\
0 & 0 & 0 & -0.455 & 0.008 & 0.038 & 0.5 & 0 & 0 & 0.143 & -0.132 & 0.061 \\
0 & 0 & 0 & -0.061 & -0.311 & 0.172 & 0 & 1.5 & 0 & -0.071 & 0.262 & 0.215 \\
0 & 0 & 0 & -0.024 & 0.134 & -0.342 & 0 & 0 & 1.5 & 0.115 & 0.222 & 0.228 \\
0.455 & 0.061 & 0.024 & 0 & 0 & 0 & 0.143 & -0.071 & 0.115 & 0.5 & 0 & 0 \\
-0.008 & 0.311 & -0.134 & 0 & 0 & 0 & -0.132 & 0.262 & 0.222 & 0 & 0.5 & 0 \\
-0.038 & -0.172 & 0.342 & 0 & 0 & 0 & 0.061 & 0.215 & 0.228 & 0 & 0 & 0.5 \\
\end{bmatrix}
\end{split}
\label{eq:feynman_re_nu_mat}
\end{equation}

\noindent The ground state of $H^{\text{Feynman}}_{Re(\nu)}$, normalized to a magnitude of $\sqrt{2}$ and with the argument of the first element set to 0, is

\begin{equation}
\ket{FCH_{Re} g.s.} = (1, 0, 0, -0.285902, 0.263185, -0.121759, 0, 0, 0, - 0.909988, 0.0167209, 0.0762265)
\label{eq:FCHRe_groundstate}
\end{equation}

\noindent Applying the mapping in Eq. \ref{eq:realtoimagmapping_sv} to $\ket{FCH_{Re} g.s.}$ in reverse returns $\ket{FCH g.s.}$ from Eq. \ref{eq:FCH_groundstate}, as expected.

In the absence of a straightforward mapping from the Feynman clock Hamiltonian to an Ising model, the Feynman clock Hamiltonian can be mapped to a QUBO problem using the Quantum Annealing Eigensolver (QAE), which was previously employed in Refs. \cite{illa2022basic, Chang2019, Teplukhin2019, Teplukhin2020, Rahman2021}. The QUBO problem matrix, $Q^{Feynman}_{QAE}$, that the Feynman clock Hamiltonian is mapped to in QAE is 

\begin{equation}
    Q^{Feynman}_{QAE} = H^{Feynman}_{Re} \otimes \begin{pmatrix}
    2^{-K + 1} 2^{-K + 1} & 2^{-K + 1} 2^{-K + 2} & \ldots & 2^{-K + 1} 2^{-1} & - 2^{-K + 1} 2^{0} \\
    2^{-K + 2} 2^{-K + 1} & 2^{-K + 2} 2^{-K + 2} & \ldots & 2^{-K + 2} 2^{-1} & - 2^{-K + 2} 2^{0} \\
    \vdots & \vdots & \ddots & \vdots & \vdots \\
    2^{-1} 2^{-K + 1} & 2^{1} 2^{-K + 2} & \ldots & 2^{-1} 2^{-1} & - 2^{0} 2^{0} \\
    -2^{0} 2^{-K + 1} & -2^{0} 2^{-K + 2} & \ldots & -2^{0} 2^{-1} &  2^{0} 2^{0} 
    \end{pmatrix}
    \label{eq:QFeynman}
\end{equation}

\noindent where K is the factor by which the dimensionality of the vector-space that $H^{Feynman}_{Re}$ acts on is increased when it is mapped to QAE.  After $Q^{Feynman}_{QAE}$ is sent to {\tt Advantage} for an anneal, {\tt Advantage} returns a result for the binary variables $q^{\alpha}_i$, as described in Sec. \ref{sec:dwaveimplementationdescription}. These binary variables are then used to obtain a result for the amplitudes $a^{\alpha}$ of the real-valued Feynman clock ground-state statevector, like so:

\begin{equation}
    a_{\alpha} = - q_{K}^{\alpha} + \sum_{i = 1}^{K - 1} \frac{q_{i}^{\alpha}}{2^{K - i}}
    \label{eq:svdigitization_app}
\end{equation}

\noindent The index $\alpha$ is used to distinguish between different amplitudes of the state-vector, while the index $i$ is used to distinguish between binary variables assigned to the same amplitude. One important detail is that in an ideal scenario, QAE's results for $a_{\alpha}$ are accurate down to an uncertainty of $2^{-K + 1}$.

QAE has the advantage of being able to obtain the full state-vector after the time evolution with a single anneal. This is important, because {\tt Advantage} exhibits a high level of noise and obtaining the full state-vector with a single anneal enables a user to mitigate the error by running multiple anneals and selecting the result with the minimal expectation value of the $H_f$ that the annealer is set to. Additionally, QAE can obtain time evolution without Trotter error \cite{illa2022basic}. However, because each amplitude in the state-vector is mapped to a register of qubits and state-vectors of states of $N_\nu$ neutrinos have $n_f^{N_\nu}$ amplitudes, the number of {\tt Advantage} qubits required to simulate an $N_\nu$-neutrino system scales as $O(n_f^{N_\nu})$, which is no improvement over the requirements of an analogous simulation on classical devices. 

Let us consider a scenario where $Q^{Feynman}_{QAE}$ is obtained for $H^{Feynman}_{\nu}$ from Eq. \ref{eq:svdigitization_app} with $K = 8$, and then the annealing process described above is done. $H^{Feynman}_{\nu}$ does not discriminate between states with the same overall phase, so several measurement results for the $q^{\alpha}_i$ values are possible. In the limit of $K \rightarrow \infty$ and no noise, all of these results will reproduce up to an overall phase the Feynman clock Hamiltonian ground state from Eqn. \ref{eq:FCHRe_groundstate} once the mapping in Eqn. \ref{eq:svdigitization_app} is applied. For simplicity's sake we will consider one possible result, shown in Tab. \ref{tab:qialpha_label}:

\begin{table}[ht]
    \centering
    \begin{tabular}{c|c c c c c c | c c c c c c}
        & \multicolumn{6}{c}{Real} & \multicolumn{6}{c}{Imaginary} \\
        & \multicolumn{3}{c}{t = 0} & \multicolumn{3}{c}{t = $10^{-13}$} & \multicolumn{3}{c}{t = 0} & \multicolumn{3}{c}{t = $10^{-13}$} \\
        & $\ket{\nu_e}$ & $\ket{\nu_\mu}$ & $\ket{\nu_\tau}$ & $\ket{\nu_e} $ & $\ket{\nu_\mu} $ & $\ket{\nu_\tau} $ & $\ket{\nu_e}$ & $\ket{\nu_\mu}$  & $\ket{\nu_\tau}$ & $\ket{\nu_e} $ & $\ket{\nu_\mu} $ & $\ket{\nu_\tau} $ \\
        $\alpha:$ & 1 & 2 & 3 & 4 & 5 & 6 & 7 & 8 & 9 & 10 & 11 & 12 \\
        \hline
        $q^{\alpha}_1$ & 1 & 0 & 0 & 1 & 0 & 1 & 0 & 0 & 0 & 0 & 0 & 0 \\
        $q^{\alpha}_2$ & 1 & 0 & 0 & 0 & 1 & 1 & 0 & 0 & 0 & 0 & 0 & 0 \\
        $q^{\alpha}_3$ & 1 & 0 & 0 & 1 & 0 & 1 & 0 & 0 & 0 & 0 & 0 & 0 \\
        $q^{\alpha}_4$ & 1 & 0 & 0 & 1 & 0 & 0 & 0 & 0 & 0 & 1 & 0 & 1 \\
        $q^{\alpha}_5$ & 1 & 0 & 0 & 0 & 0 & 0 & 0 & 0 & 0 & 0 & 0 & 0 \\
        $q^{\alpha}_6$ & 1 & 0 & 0 & 1 & 0 & 0 & 0 & 0 & 0 & 1 & 1 & 0 \\
        $q^{\alpha}_7$ & 1 & 0 & 0 & 1 & 1 & 0 & 0 & 0 & 0 & 1 & 0 & 1 \\
        $q^{\alpha}_8$ & 0 & 0 & 0 & 1 & 0 & 1 & 0 & 0 & 0 & 1 & 0 & 0 \\
    \end{tabular}
    \caption{The binary variable results from submitting $Q^{Feynman}_{QAE}$ for $H^{Feynman}_{\nu}$ from Eq. \ref{eq:svdigitization_app} with $K = 8$ to {\tt Advantage}. Each column lists out the results for all of the binary variables assigned to one amplitude within the state-vector of the ground state of $H^{Feynman}_{Re(\nu)}$. The top row lists out the indices $\alpha$ for each of the aforementioned amplitudes, and above the top row the neutrino flavor and time that each $\alpha$ corresponds to and whether each $\alpha$ corresponds to a real or imaginary component of $H^{Feynman}_{\nu}$'s ground state are listed. }
    \label{tab:qialpha_label}
\end{table}

\noindent Applying the mapping in Eqn. \ref{eq:svdigitization_app} to the results in Tab. \ref{tab:qialpha_label} gives the statevector

\begin{equation}
    \vec{a} = (0.992188, 0, 0, -0.289063, 0.257813, -0.125, 0, 0, 0, -0.914063, 0.015625, 0.0703125)
    \label{eq:gottenstatevector}
\end{equation}

\noindent which agrees with $\ket{FCH_{Re} g.s.}$ from Eq. \ref{eq:FCHRe_groundstate} to within $\frac{1}{2^7}$, as expected.



{\tt Advantage}'s hardware runs on an architecture known as Pegasus. This means that the two-qubit couplings $J_{ij}$ can act between any qubit and one of the 15 other qubits on {\tt Advantage} that Pegasus connects it to \mbox{\cite{mcgeoch2022advantage}}. However, collective neutrino oscillations require coupling between all pairs of simulated neutrinos, which combined with the dense matrices that come out of the QUBO problem mapping in Eq. \ref{eq:QFeynman} generally means that that all-to-all coupling is effectively required to solve neutrino dynamics problems on {\tt Advantage}. The required all-to-all coupling is obtained on {\tt Advantage} through a process called minor-embedding. Minor-embedding works by mapping each logical qubit in a given annealing problem to a set of physical qubits called a ``chain''. All qubits in a chain are fixed to the same value by setting the $Q_{ij}$ parameters between them to a negative value high enough in magnitude (which is called the ``chain-strength'') to do so and that is set by the user \mbox{\cite{dwavesysdocs}}. In this project, the chain-strength is set to 1, because it is high enough to prevent almost all differences in value between qubits in a chain (``chain-breaks''), and higher values have been found to interfere with convergence to the final ground state. Because all of the qubits in a chain are the same value, the chains can be treated as 1 qubit with a connectivity equal to the sum of the connectivities of the qubits minus the number of connections between qubits on the same chain. In this project, D-Wave's provided minor-embedders, which have a quadratic complexity \mbox{\cite{dwavesysdocs}}, are used for this purpose.

{\tt Advantage}'s 5000+ qubits are insufficient to run the problems in this project with a $K$ sufficiently high to achieve the precisions achieved in this project. This is especially true after accounting for the extra overhead from applying minor-embedding. Thus, instead of QAE, AQAE (Adaptive Quantum Annealing Eigensolver) is used instead. This method has been used before in Refs. \cite{Chang2019, Rahman2021, illa2022basic}. The procedure for the method is decribed in Sec. \ref{sec:timeevontoQUBOmapping}, but in this appendix its derivation is described.

The primary step of deriving AQAE is adapting the QUBO problem mapping in Eqn. \ref{eq:QFeynman} so that the binary variable results returned at the end of the anneal encodes not the Feynman clock Hamiltonian ground state, but an update to a prior result for the ground state. This step is taken directly from Appendix A of Ref. \cite{illa2022basic}. This adaptation starts with the stipulation that Eq. \ref{eq:svdigitization_app} must adapt so that an annealing run with a given value of the integer variable $z$ takes in a prior $a^{(z)}$ for the ground state of $H^{Feynman}_{Re}$ and, using the binary variable results $q_i^{\alpha}$ from the anneal, obtain an updated ground state result $a^{(z + 1)}$ like so: 

\begin{equation}
        a^{(z + 1)}_\alpha = a^{(z)}_\alpha - 2^{-z} q^{\alpha}_K + \sum_{i = 1}^{K - 1} \frac{q^{\alpha}_i}{2^{K - i -z}}
        \label{eq:svdigitization_basicaqae_app}
\end{equation}

\noindent Taking advantage of Eqn. \ref{eq:svdigitization_basicaqae_app},

\begin{equation}
\begin{split}
    a^{(z + 1)}_\alpha (H^{Feynman}_{Re})_{\alpha \beta}  a^{(z + 1)}_\beta = [a^{(z)}_\alpha - 2^{-z} q^{\alpha}_K + \sum_{i = 1}^{K - 1} \frac{q^{\alpha}_i}{2^{K - i -z}}] (H^{Feynman}_{Re})_{\alpha \beta} [a^{(z)}_\beta - 2^{-z} q^{\beta}_K + \sum_{i = 1}^{K - 1} \frac{q^{\beta}_i}{2^{K - i -z}}] \\
    = a^{(z)}_\alpha (H^{Feynman}_{Re})_{\alpha \beta}  a^{(z)}_\beta + 2 a^{(z)}_\alpha (H^{Feynman}_{Re})_{\alpha \beta} [- 2^{-z} q^{\beta}_K + \sum_{i = 1}^{K - 1} \frac{q^{\beta}_i}{2^{K - i -z}}] + \\
     [- 2^{-z} q^{\alpha}_K + \sum_{i = 1}^{K - 1} \frac{q^{\alpha}_i}{2^{K - i -z}}] (H^{Feynman}_{Re})_{\alpha \beta} [- 2^{-z} q^{\beta}_K + \sum_{i = 1}^{K - 1} \frac{q^{\beta}_i}{2^{K - i -z}}] 
     \end{split}
\label{eq:equivalentmatrixelements_app}
\end{equation}

\noindent and the minimization of the left-hand side is equivalent to the minimization of the right-hand side of Eq. \ref{eq:equivalentmatrixelements_app}. The first term of the right-hand side is immutable by the results of an anneal, since it is only dependent on $H^{Feynman}_{Re}$ and the prior on its ground state. The third term of the right-hand side reproduces the mapping in Eqn. \ref{eq:QFeynman}. To turn the second term into a prescription for the mapping to a QUBO problem, one can take advantage of the fact that since the binary variables $q^{\beta}_i$ are restricted to the values 0 and 1, $q^{\beta}_i q^{\beta}_i$ = $q^{\beta}_i$. Combining the results of the second and third terms' implications for the QUBO problem gives

\begin{equation}
\begin{split}
    Q^{Feynman}_{AQAE} = H^{Feynman}_{Re} \otimes \begin{pmatrix}
    2^{-K + 1} 2^{-K + 1} & 2^{-K + 1} 2^{-K + 2} & \ldots & 2^{-K + 1} 2^{-1} & - 2^{-K + 1} 2^{0} \\
    2^{-K + 2} 2^{-K + 1} & 2^{-K + 2} 2^{-K + 2} & \ldots & 2^{-K + 2} 2^{-1} & - 2^{-K + 2} 2^{0} \\
    \vdots & \vdots & \ddots & \vdots & \vdots \\
    2^{-1} 2^{-K + 1} & 2^{1} 2^{-K + 2} & \ldots & 2^{-1} 2^{-1} & - 2^{0} 2^{0} \\
    -2^{0} 2^{-K + 1} & -2^{0} 2^{-K + 2} & \ldots & -2^{0} 2^{-1} &  2^{0} 2^{0} 
    \end{pmatrix} + \\
    diag(a^{(z)} H^{Feynman}_{Re}) \otimes
diag(2^{-K + 1 - z}, 2^{-K + 2 - z}, \ldots , 2^{1-z} , -2^{-z})
\end{split}
\label{eq:QFeynman_AQAE_app}
\end{equation}

Starting with a prior $a^{(0)}$ for $H^{Feynman}_{Re}$'s ground state; updating it by using Eq. \ref{eq:QFeynman_AQAE_app} to obtain a QUBO problem, having {\tt Advantage} obtain the QUBO problems' $q_i^{\alpha}$ binary variable results, and finally using Eq. \ref{eq:svdigitization_basicaqae_app}; and repeating the update with $z$ starting with 0 and increasing by 1 with each repetition until convergence gives the implementation of AQAE in Ref. \cite{illa2022basic}. To obtain the process in Sec. \ref{sec:timeevontoQUBOmapping}, several more changes were made to this process. The first of these modifications is to add a second update right after each update. The second update is identical in all respects to the original update, with the exception that its QUBO problem mapping is 

\begin{equation}
\begin{split}
    Q^{Feynman}_{AQAEv2} = H^{Feynman}_{Re} \otimes \begin{pmatrix}
    2^{-K + 1} 2^{-K + 1} & 2^{-K + 1} 2^{-K + 2} & \ldots & 2^{-K + 1} 2^{-1} & - 2^{-K + 1} 2^{0} \\
    2^{-K + 2} 2^{-K + 1} & 2^{-K + 2} 2^{-K + 2} & \ldots & 2^{-K + 2} 2^{-1} & - 2^{-K + 2} 2^{0} \\
    \vdots & \vdots & \ddots & \vdots & \vdots \\
    2^{-1} 2^{-K + 1} & 2^{1} 2^{-K + 2} & \ldots & 2^{-1} 2^{-1} & - 2^{0} 2^{0} \\
    -2^{0} 2^{-K + 1} & -2^{0} 2^{-K + 2} & \ldots & -2^{0} 2^{-1} &  2^{0} 2^{0} 
    \end{pmatrix} - \\
    diag(a^{(z)} H^{Feynman}_{Re}) \otimes
diag(2^{-K + 1 - z}, 2^{-K + 2 - z}, \ldots , 2^{1-z} , -2^{-z})
\end{split}
\label{eq:QFeynman2_AQAE_app}
\end{equation}

\noindent After this modification, the result of applying the original update to the prior $a^{z}$ is 

\begin{equation}
        b^{(z)}_\alpha = a^{(z)}_\alpha - 2^{-z} q^{\alpha}_K + \sum_{i = 1}^{K - 1} \frac{q^{\alpha}_i}{2^{K - i -z}}
        \label{eq:svdigitization_dualaqaepart1_app}
\end{equation}

\noindent and the result of applying the second update to $b^{(z)}_\alpha$, the result after the original update, is 

\begin{equation}
        a^{(z + 1)}_\alpha = b^{(z)}_\alpha + 2^{-z} q^{\alpha}_K - \sum_{i = 1}^{K - 1} \frac{q^{\alpha}_i}{2^{K - i -z}}
        \label{eq:svdigitization_dualaqaepart2_app}
\end{equation}

\begin{figure}
    \centering
    \includegraphics[width=0.49\linewidth]{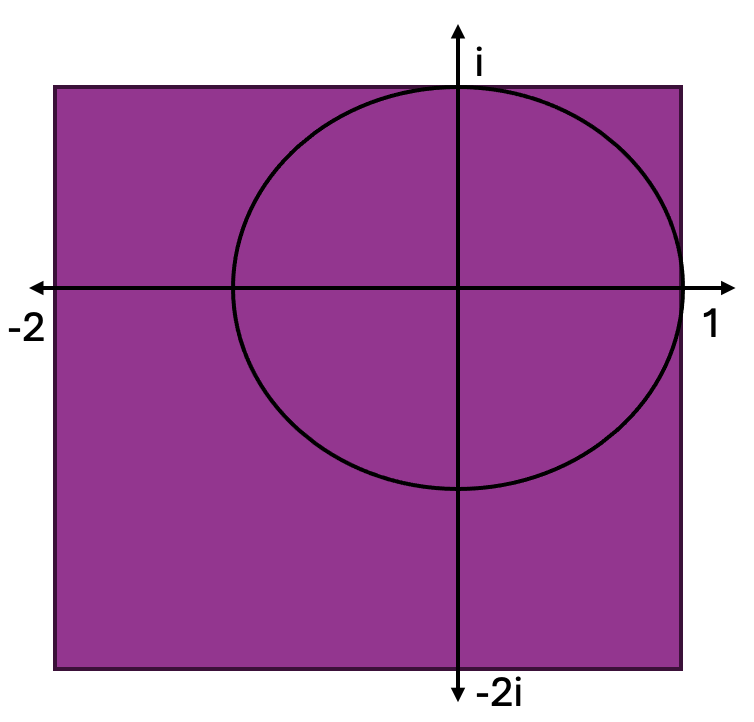}  
    \includegraphics[width=0.49\linewidth]{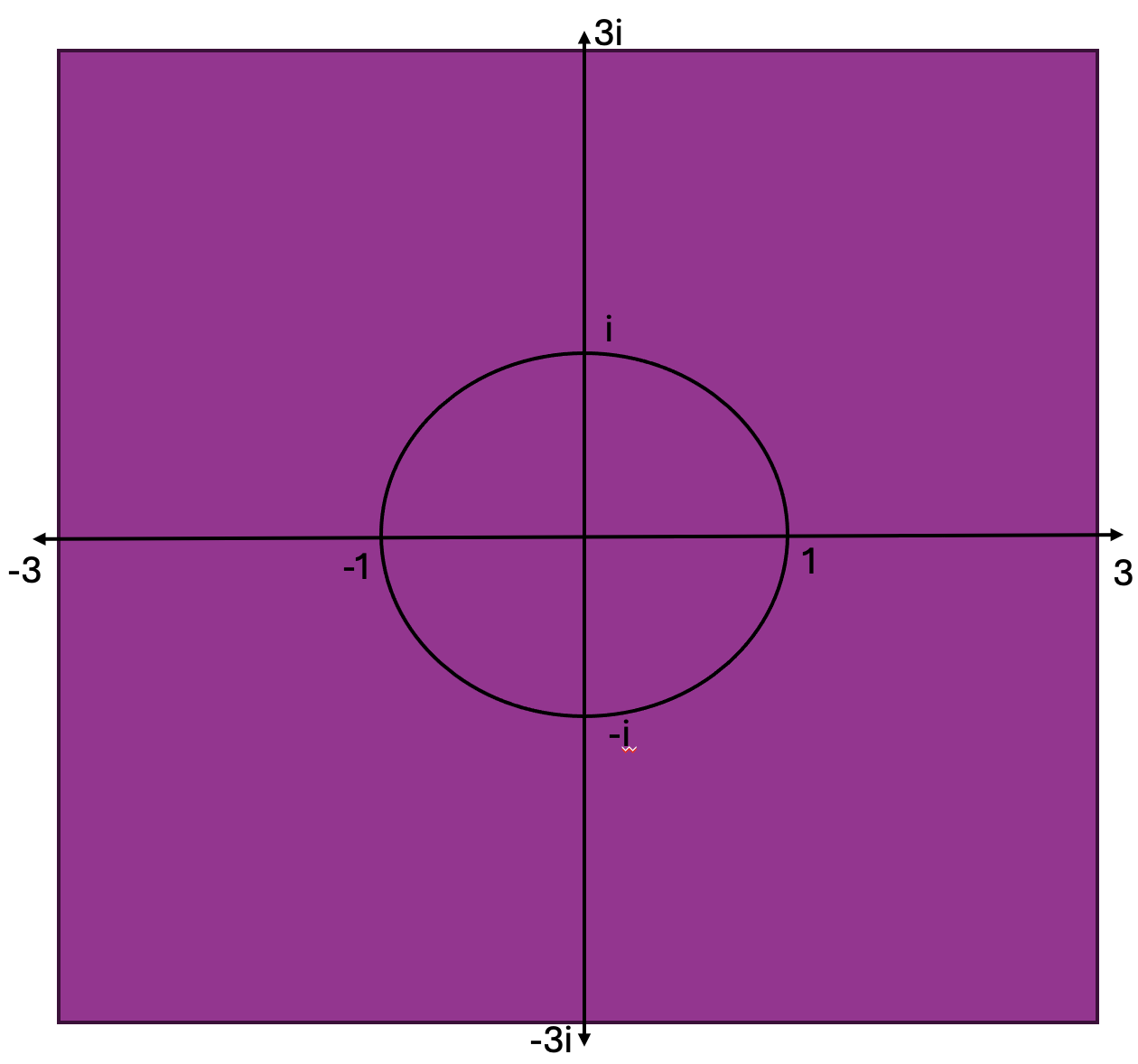}  \\
    \includegraphics[width=0.49\linewidth]{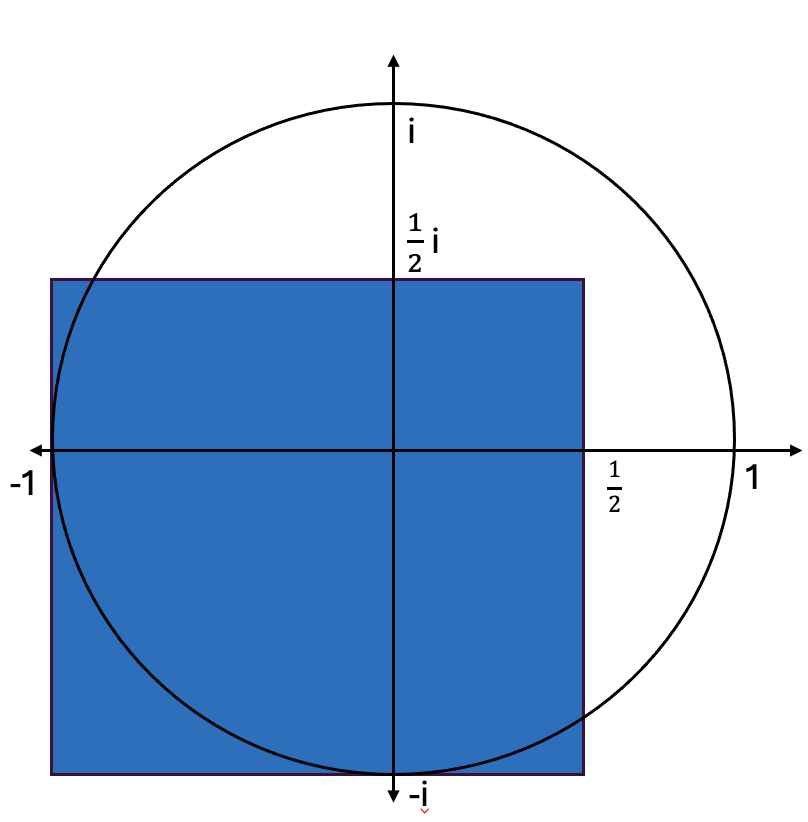}  
    \includegraphics[width=0.49\linewidth]{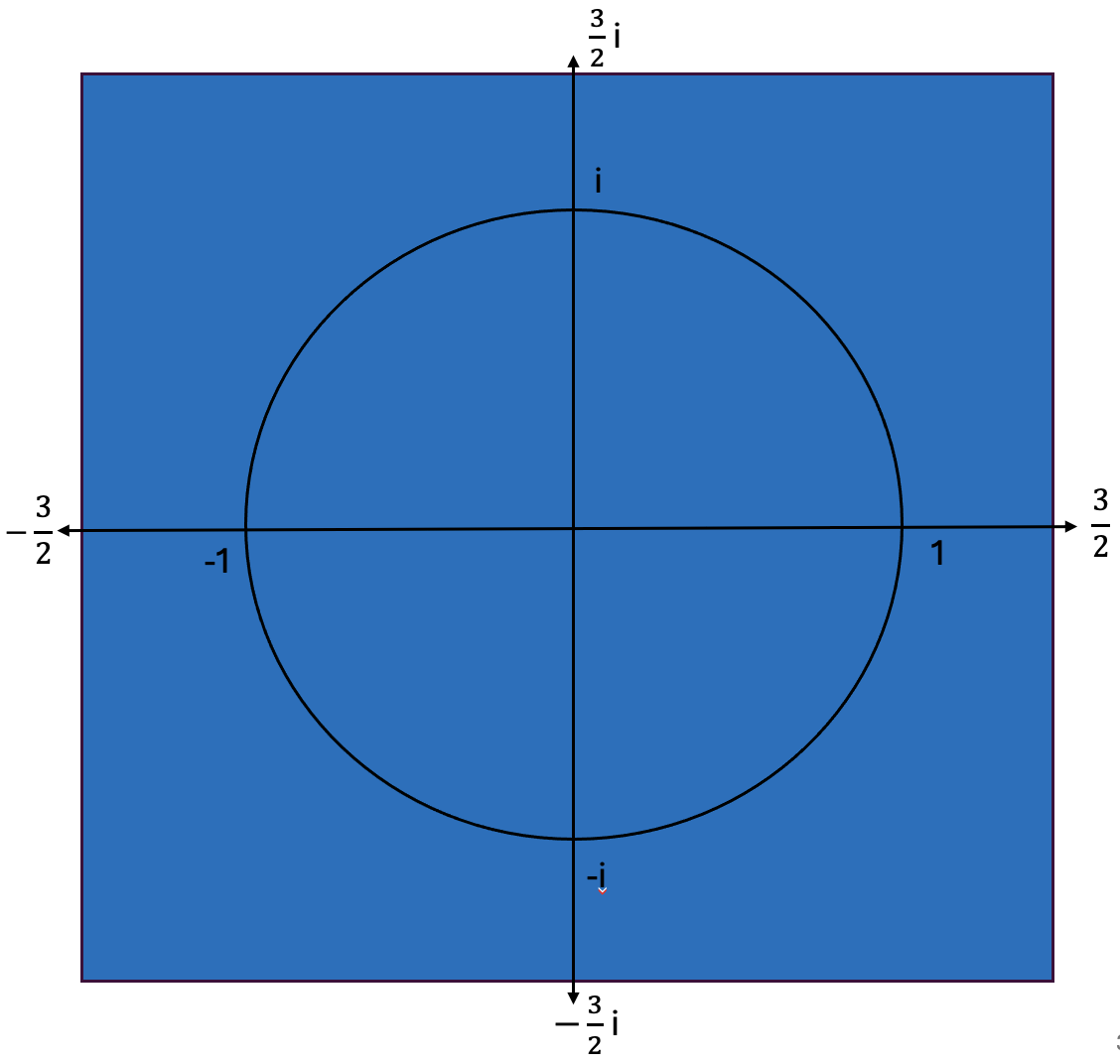}  \\
    \caption{ Color-shaded regions represent possible values of a given amplitude $a^{(z)}_{\alpha}$ within the state-vector for the ground state of the Feynman clock Hamiltonian in the limit of $z \rightarrow \infty$ given that $a^{(z_0)}_{\alpha}$ = 0 and K = 1.  Top: $z_0$ = 0; Bottom: $z_0$ = 1. Left: the case in which AQAE is done as implemented in Ref. \cite{illa2022basic}; Right: the case in which AQAE is done with the near-duplicate second update defined by Eqs. \ref{eq:QFeynman2_AQAE_app} and \ref{eq:svdigitization_dualaqaepart2_app} included on top of the procedure from Ref. \cite{illa2022basic}.}
    \label{fig:AQAE_amplitudevalueranges_app}
\end{figure}

\noindent The motivation behind adding the second update is robustness against noise. As seen in Fig. \ref{fig:AQAE_amplitudevalueranges_app}, the full range of physical values of a given amplitude in the state-vector for the ground state of the Feynman clock Hamiltonian is accessible if one starts at $z$= 0 but not all of it accessible if one starts at $z$= 1 if one implements AQAE as it is implemented in Ref. \cite{illa2022basic} with K = 2. This means that if there is a bit-flip error, which are common on {\tt Advantage}, in the $z=0$ iteration that causes the output of the $z=0$ update to be wrong, one may never converge to the correct result. The near-duplicate second update mitigates this problem, since both the second update and (as seen on the right-hand side of Fig. \ref{fig:AQAE_amplitudevalueranges_app}) the updates with the next z-value can correct for the bit-flip error.  Incidentally, the second update also enables K = 1 to converge, which Ref. \cite{illa2022basic} failed to achieve.

Second, the prior $a^{(0)}$ on the Feynman clock Hamiltonian is set to $\ket{\Psi_0} \otimes \ket{t_i} + \ket{0} \otimes \ket{t_f}$, where $\ket{t_i}$ is the initial-time state, $\ket{t_f}$ is the final-time state, $\ket{\Psi_0}$ is the system's initial state, and $\ket{0}$ is the state with the same vector-space size as $\ket{\Psi_0}$ with a norm of 0. This is done since because $\ket{\Psi_0}$ is already known, including it in the prior $a^{(0)}$ will help speed up convergence, especially on a noisy device such as {\tt Advantage}. The final state is priored on the norm-0 state on the principle of starting with a uniform prior for an unknown state, as the norm-0 state is the state from which all of the physical final states are equidistant.

Third, the penalty term $C_0$ from Eq. \ref{eq:feynmanclockham} is set to 0 and instead its function is taken up by adding a penalty term $Q_0$ to the QUBO problem matrix,

\begin{equation}
    Q_0 = diag(1, 0, 1, 0) \otimes 3.5 \mathbb{I}_{Kn, Kn}
    \label{eq:QUBOpenaltyterm_app}
\end{equation}

\noindent This is because since the initial state of the system is already known and already implemented in the prior $a^{(0)}$, all of the subsequent updates to the initial ground state should be 0. Hence, the function of the penalty term should be to fix the qubit-measurements accordingly and to do so with the highest possible penalty for deviations in order to drown out the ever-present noise on {\tt Advantage}. This means that all qubits whose measurements produce binary variables which update the initial state should be in the state that produces a measurement of $q_i^{\alpha} = 0$, and one way to do that is to set the diagonal terms of the QUBO problem matrix that act on said binary variables to the maximum value possible \cite{dwavesysdocs}. Given that diagonal elements on a QUBO matrix sent to {\tt Advantage } have a maximal absolute value of 4 and if any elements exceed this absolute value the whole matrix is scaled down \cite{dwavesysdocs}, 3.5 is chosen in order to provide an allowance for the other terms that contribute to the QUBO matrix.

Fourth, small QUBO matrix elements are liable to be drowned out by noise, and the QUBO matrix elements in Eqns. \ref{eq:QFeynman_AQAE_app} and \ref{eq:QFeynman2_AQAE_app} become smaller as $z$ increases. To prevent noise from overwhelming measurements at high $z$, an extra prefactor of $2^{2z - 1}$ is multiplied into all terms contributing to the QUBO matrix except the penalty term.

\section{Benchmarking}
\label{app:nealbenchmarking}

\begin{figure}[ht]

    \raggedleft
    \includegraphics[width=0.05\textwidth]{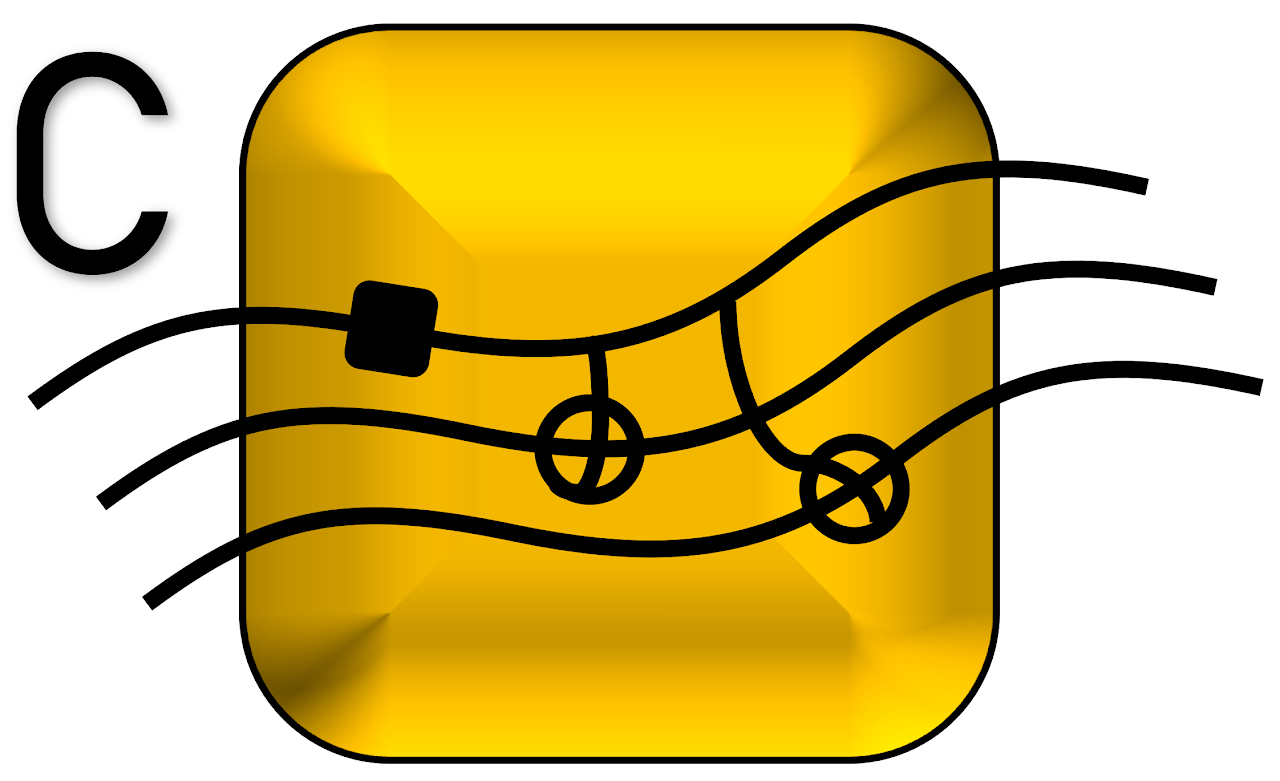}
    
    \centering

    \includegraphics[width=0.32\textwidth]{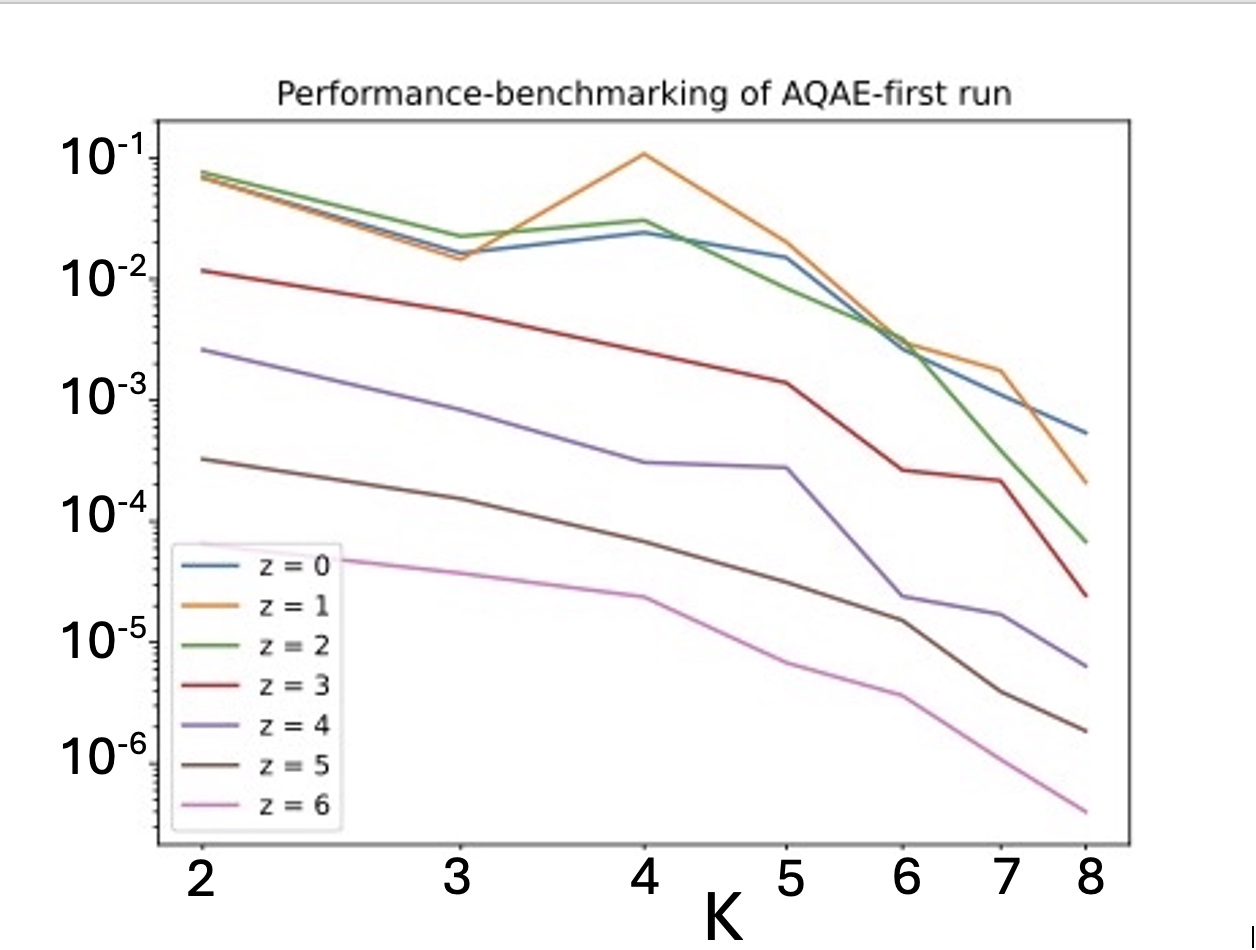}
    \includegraphics[width=0.32\textwidth]{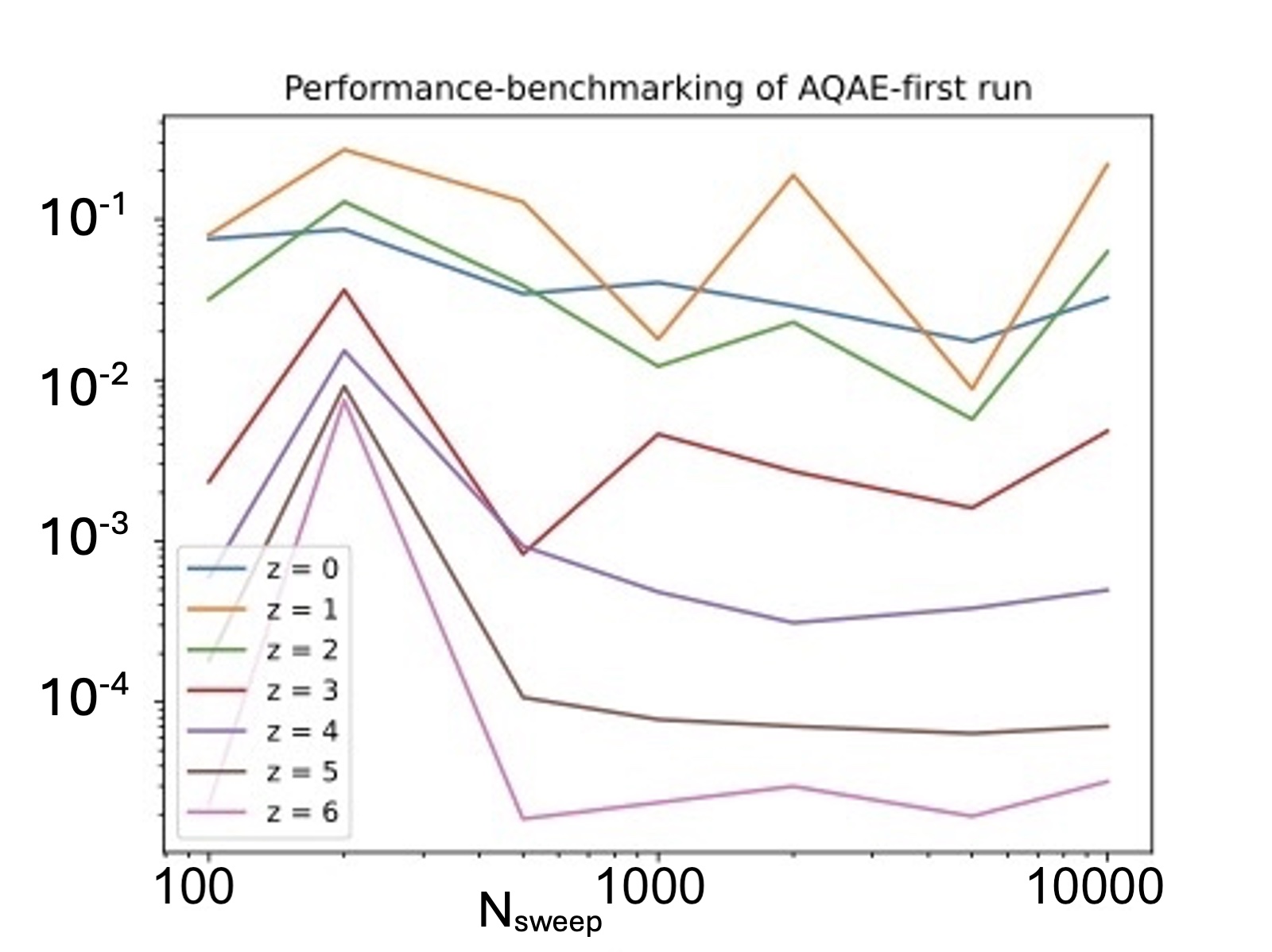}
    \includegraphics[width=0.32\textwidth]{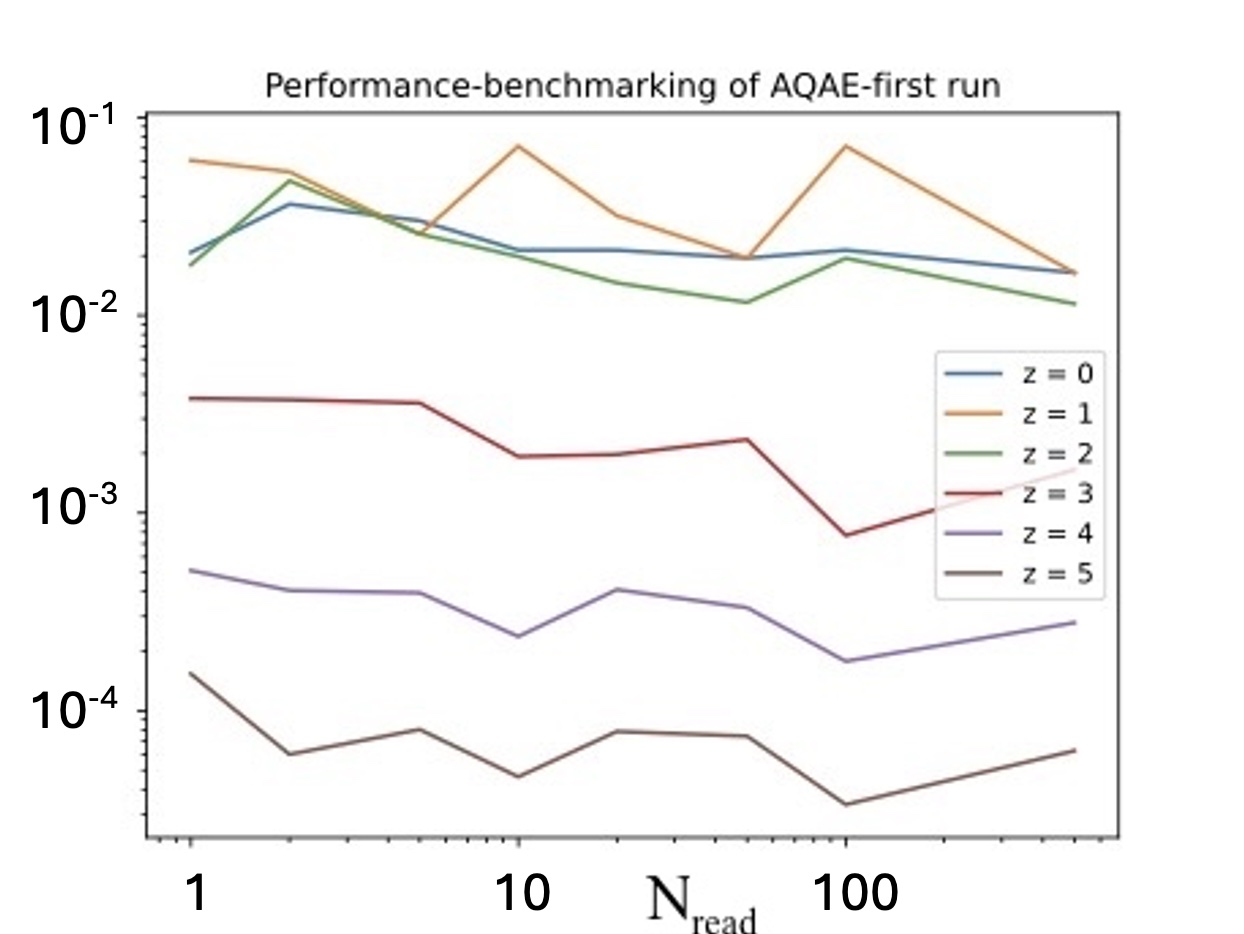} 

    \caption{From left to right, the effect of z and K, of z and the number of sweeps in an anneal (which is a surrogate for anneal-time), and of z and the number of anneal-repetitions per call to the annealer on 1 minus the dot product of the final state-vector of the time-evolution of 2 neutrinos as calculated by Neal and the same state-vector as calculated using exact linear algebra. The initial state is $\ket{\nu_e \nu_\mu}$, the Hamiltonian used for the evolution is Equation \ref{eq:H_collectiveneutrinooscilation}, the parameters are found in Table \ref{tab:2flavexactsimulationpars}, and the time-interval evolved over is $10^{12}$ $eV^{-1}$ . Each entry is the lowest-energy final result out of all the anneal-repetitions in its AQAE step. The yellow icon at the top-right indicates a noisy classical simulation \cite{Klco2020}.}
    \label{fig:nealbench}
\end{figure}

AQAE's performance depends on the number of repetitions of the anneal, the annealing time, and the constants K and z in Eq. \ref{eq:QFeynman_AQAEforwardorder},  \ref{eq:azposteriorfindingnormaldig}, \ref{eq:QFeynman_AQAEreverseorder}, and \ref{eq:azposteriorfindingreversedig} which determine the digitization and the number of zoom-steps. Thus, the first step before running any process on the annealer is to use D-Wave's simulated classical thermal annealer, \texttt{neal}, to assess the best way to use these parameters to optimize device performance. The Hamiltonian in Eq. \ref{eq:H_collectiveneutrinooscilation} was used in this case, with N = 2. The parameters are the same ones in Tab. \ref{tab:2flavexactsimulationpars}.
$\vec{B}(E)$ was simply taken from Ref. \cite{siwach2023entanglement}: $\vec{B}(E) = (0, 0, -\frac{\delta m^2}{4 E}, 0, 0, 0, 0, -\frac{\Delta m^2}{4 E})$. The starting state was $\ket{\nu_e \nu_\mu}$. The time-increment evolved over was t = $10^{12} eV^{-1}$. The reverse-digitization process and the fixing of the penalty term $h_{i}$ values discussed in Section \ref{sec:dwaveimplementationdescription} and Appendix \ref{app:AQAEderivdemonstration} were not used; instead, the digitization procedure and the penalty term were found in the same manner as in Ref. \cite{illa2022basic}. In doing so, I replicate Ref. \cite{illa2022basic}'s result that increasing z produces a much greater improvement in precision than any of the above options, as seen in Fig. \ref{fig:nealbench}. 

\begin{figure}[ht]
    \raggedleft
    \includegraphics[width=0.05\textwidth]{quantum-simulation-quantum-hardware.png}

    \centering
    \includegraphics[width=0.50\textwidth]{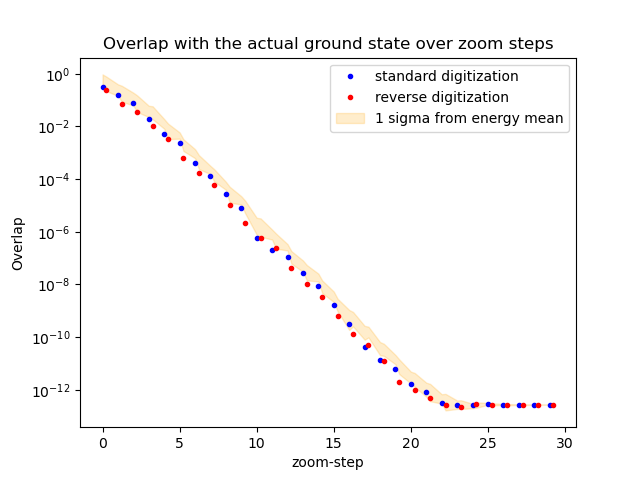}
    \caption{The overlap between results of time-evolution with AQAE of a three-flavor, two-neutrino system starting as one electron and one muon neutrino and evolved over a time-increment of $10^{10} eV^{-1}$  and the actual final state. $N_{reads}$ = 1000, $a_t$ = 20 $\mu s$, K = 1, and chain-strength = 1.  The blue icon at the top-right indicates simulation on a quantum device \cite{Klco2020}.}
    \label{fig:forcedpenaltyAQAEalternatingdigresults}
\end{figure}

Thus, K, anneal-time, and the number of repetitions of each anneal-step only need to be adequate to converge, and once that is assured, the way to improve precision is to increase z. The next step is to assess how many zoom-steps are needed in order to converge to a satisfactory precision. This was done by running the same simulation on {\tt Advantage}, with the reverse-digitization and fixing of the penalty term $h_{i}$ values included. Annealer parameters were K = 2, anneal-time 5 $\mu s$, 1000 repetitions of each zoom-step, and a chain-strength of 1. As seen in Fig. \ref{fig:forcedpenaltyAQAEalternatingdigresults}, it takes about 26 zoom-steps to converge.

\newpage

\section{Tables of N = 4 $\nu$-$\nu$ results for entanglement witnesses}
\label{app:n4entanglementtabs}

\begin{table}[ht]
\centering

\label{tab:N4nunuexactresults_S3}
\begin{tabular}{c|c|c|c|c}
\hline
time ($eV^{-1}$) & $S_3(t)$ \\
\hline
$1.1 \times 10^{12}$ & $0.222927229(^{+619}_{-1527})$ \\
$2.2 \times 10^{12}$ & $0.623264640(^{+50}_{-3222})$ \\
$3.3 \times 10^{12}$ & $1.0340534849(^{+1546}_{-660})$ \\
$4.4 \times 10^{12}$ & $1.3420253885(^{+18009}_{-6643})$ \\
$5.5 \times 10^{12}$ & $1.5060393872(^{+2875}_{-67})$ \\
$6.6 \times 10^{12}$ & $1.5718841229(^{+252}_{-308})$ \\
$7.7 \times 10^{12}$ & $1.5693430609(^{+6021}_{-1011})$ \\
$8.8 \times 10^{12}$ & $1.4631204675(^{+109}_{-6577})$ \\
$9.9 \times 10^{12}$ & $1.2671452598(^{+451}_{-54})$ \\

\end{tabular}

\caption{A table of the D-Wave {\tt Advantage} quantum annealer results for entanglement entropy for the third neutrino of the N = 4, $n_f$ = 3 system results from Fig. \ref{fig:fourneutrinodwavetoclassicalcomparison}. Error bars are 1$\sigma$ confidence intervals.}
\end{table}

\begin{table}[ht]
\centering

\label{tab:N4nunuexactresults_N3}
\begin{tabular}{c|c|c|c}
\hline
time ($eV^{-1}$) & $N_{13}(t)$ & $N_{23}(t)$ & $N_{34}(t)$ \\
\hline
$1.1 \times 10^{12}$ & $0.3720986223(^{+10285}_{-17470})$ & $0.0842779937(^{+2064}_{-9353})$ & $0.0984672558(^{+4280}_{-16009})$ \\
$2.2 \times 10^{12}$ & $0.4868408929(^{+104}_{-13601})$ & $0.1291148436(^{+194}_{-692})$ & $0.2787261578(^{+144}_{-4655})$ \\
$3.3 \times 10^{12}$ & $0.5484990103(^{+712}_{-7655})$ & $0.2091610157(^{+20706}_{-110})$ & $0.4936497469(^{+94}_{-18696})$ \\
$4.4 \times 10^{12}$ & $0.5918213991(^{+694}_{-5279})$ & $0.3460545631(^{+2102}_{-12006})$ & $0.5329880752(^{+7887}_{-255})$ \\
$5.5 \times 10^{12}$ & $0.5197404555(^{+117}_{-1131})$ & $0.4497961383(^{+364}_{-654})$ & $0.5347131445(^{+495}_{-387})$ \\
$6.6 \times 10^{12}$ & $0.4301304886(^{+1428}_{-64})$ & $0.4458340852(^{+35}_{-3029})$ & $0.6025444696(^{+695}_{-33})$ \\
$7.7 \times 10^{12}$ & $0.4824124333(^{+18476}_{-461})$ & $0.3611637035(^{+388}_{-22517})$ & $0.5514675837(^{+731}_{-19224})$ \\
$8.8 \times 10^{12}$ & $0.5822152101(^{+371}_{-1104})$ & $0.2461251687(^{+351}_{-5615})$ & $0.3860333315(^{+438}_{-1888})$ \\
$9.9 \times 10^{12}$ & $0.4597905236(^{+16}_{-956})$ & $0.1754865412(^{+301}_{-18})$ & $0.2120581628(^{+1237}_{-18})$ \\

\end{tabular}

\caption{A table of the D-Wave {\tt Advantage} quantum annealer results for negativities for the third neutrino of the N = 4, $n_f$ = 3 system results from Fig. \ref{fig:fourneutrinodwavetoclassicalcomparison}. Error bars are 1$\sigma$ confidence intervals.}
\end{table}


\newpage

\bibliography{dwave3flavpaper.bib}


\end{document}